\documentclass[11pt,a4paper]{article}

\usepackage[a4paper, left=1in, right=1in, top=1in, bottom=1in, includehead, includefoot, foot = 0pt, head=0pt]{geometry} 
\usepackage[group-separator={,},group-digits=integer]{siunitx} 
\usepackage{booktabs} 
\usepackage{enumitem} 

\usepackage[round,colon,authoryear]{natbib} 
\usepackage[pdftex,colorlinks,citecolor=blue,urlcolor=blue,linkcolor=red]{hyperref} 
\usepackage{graphicx} 
\parskip=\medskipamount 
\usepackage[small,bf,hang]{caption}	
\usepackage{amssymb} 
\usepackage{amsmath} 
\usepackage{bbm} 
\usepackage{amsbsy} 
\usepackage{mathrsfs} 
\usepackage[noend]{algorithm2e} 
\SetAlCapFnt{\small} 
\SetAlCapNameFnt{\small} 
\usepackage{multirow} 
\usepackage[dvipsnames,table]{xcolor} 
\usepackage{subcaption} 
\usepackage[hang,flushmargin]{footmisc} 
\usepackage{tikz} 
\usetikzlibrary{matrix, fit} 
\usepackage{rotfloat}
\usepackage{titling} 
\makeatletter
\renewcommand{\paragraph}{%
	\@startsection{paragraph}{4}%
	{\z@}{0.5ex \@plus 1ex \@minus .2ex}{-1em}
	{\normalfont\normalsize\bfseries}%
}
\makeatother

\usepackage{authblk} 

\usepackage{fancyhdr}
\begin{document}
\sloppy

\title{\LARGE Boosting insights in insurance tariff plans \\ with tree-based machine learning methods}
\author[2,4]{Roel Henckaerts}
\author[1]{Marie-Pier C\^{o}t\'{e}}
\author[2,3,4]{Katrien Antonio}
\author[2,4]{Roel Verbelen}
\affil[1]{\'{E}cole d'actuariat, Universit\'{e} Laval, Canada.}
\affil[2]{Faculty of Economics and Business, KU Leuven, Belgium.}
\affil[3]{Faculty of Economics and Business, University of Amsterdam, The Netherlands.}
\affil[4]{LRisk, Leuven Research Center on Insurance and Financial Risk Analysis, KU Leuven, Belgium.}
\predate{}
\postdate{}
\date{}
\maketitle
\thispagestyle{empty}

\begin{abstract}
\noindent Pricing actuaries typically operate within the framework of generalized linear models (GLMs). With the upswing of data analytics, our study puts focus on machine learning methods to develop full tariff plans built from both the frequency and severity of claims. We adapt the loss functions used in the algorithms such that the specific characteristics of insurance data are carefully incorporated: highly unbalanced count data with excess zeros and varying exposure on the frequency side combined with scarce, but potentially long-tailed data on the severity side. A key requirement is the need for transparent and interpretable pricing models which are easily explainable to all stakeholders. We therefore focus on machine learning with decision trees: starting from simple regression trees, we work towards more advanced ensembles such as random forests and boosted trees. We show how to choose the optimal tuning parameters for these models in an elaborate cross-validation scheme, we present visualization tools to obtain insights from the resulting models and the economic value of these new modeling approaches is evaluated. Boosted trees outperform the classical GLMs, allowing the insurer to form profitable portfolios and to guard against potential adverse risk selection. \\ [2mm]
\end{abstract}

\section{Introduction}
Insurance companies bring security to society by offering protection against financial losses. They allow individuals to trade uncertainty for certainty, by transferring the risk to the insurer in exchange for a fixed premium. An insurer sets the price for an insurance policy before its actual cost is revealed. Due to this phenomenon, known as the reverse production cycle of the insurance business, it is of vital importance that an insurer properly assesses the risks in its portfolio. To this end, tools from predictive modeling come in handy.

The insurance business is highly data driven and partly relies on algorithms for decision making. In order to price a contract, property and casualty (P\&C, or: general, non-life) insurers predict the loss cost $y$ for each policyholder based on his/her observable characteristics~$\boldsymbol{x}$. The insurer therefore develops a predictive model~$f$, mapping the risk factors~$\boldsymbol{x}$ to the predicted loss cost~$\hat{y}$ by setting $\hat{y}=f(\boldsymbol{x})$. For simplicity, this predictive model is usually built in two stages by considering separately the frequency and severity of the claims. Generalized linear models (GLMs), introduced by \citet{Nelder1972}, are the industry standard to develop state-of-the-art analytic insurance pricing models \citep{Haberman1996,Jong2008,Frees2015}. Pricing actuaries often code all risk factors in a categorical format, either based on expert opinions  \citep{Frees2008,Antonio2010} or in a data-driven way \citep{Henckaerts2018}. GLMs involving only categorical risk factors result in predictions available in tabular format, that can easily be translated into interpretable tariff plans.

Technological advancements have boosted the popularity of machine learning and big data analytics, thereby changing the landscape of predictive modeling in many business applications. However, few papers in the insurance literature go beyond the actuarial comfort zone of GLMs. \citet{Dal2010} contrasts the performance of various machine learning techniques to predict claim frequency in the Allstate Kaggle competition. \citet{Guelman2012} compares GLMs and gradient boosted trees for predicting the accident loss cost of auto at-fault claims. \citet{Liu2014} approach the claim frequency prediction problem using multi-class AdaBoost trees. \citet{Wuthrich2019} and \citet{Zochbauer2017} show how tree-based machine learning techniques can be adapted to model claim frequencies. \citet{Yang2018} predict insurance premiums by applying a gradient boosted tree algorithm to Tweedie models. \citet{Pesantez2019} employ XGBoost to predict the occurrence of claims using telematics data. \citet{Ferrario2018} and \citet{Schelldorfer2019} propose neural networks to model claim frequency, either directly or via a nested GLM. Machine learning techniques have also been used in other insurance applications, such as policy conversion or customer retention \citep{Spedicato2018}, renewal pricing \citep{Krasheninnikova2019} and claim fraud detection \citep{Wang2018}.

Insurance pricing models are heavily regulated and they must meet specific requirements before being deployed in practice, posing some challenges for machine learning algorithms. Firstly, the European Union's General Data Protection Regulation \citep{GDPR2018}, effective May 25, 2018, establishes a regime of ``algorithmic accountability" of decision-making machine algorithms. By law, individuals have the right to an explanation of the logic behind the decision \citep{Kaminski2018}, which means that pricing models must be transparent and easy to communicate to all stakeholders. Qualified transparency \citep{Pasquale2015} implies that customers, managers and the regulator should receive information in different degrees of scope and depth. Secondly, every policyholder should be charged a fair premium, related to his/her risk profile, to minimize the potential for adverse selection \citep{Dionne1999}. If the heterogeneity in the portfolio is not carefully reflected in the pricing, the good risks will be prompt to lapse and accept a lower premium elsewhere, leaving the insurer with an inadequately priced portfolio. Thirdly, the insurer has the social role of creating solidarity among the policyholders. The use of machine learning for pricing should in no way lead to an extreme ``personalization of risk'' or discrimination, e.g.,~in the form of extremely high premiums \citep{Oneil2017} for some risk profiles that actually entail no risk transfer. By finding a trade-off between customer segmentation and risk pooling, the insurer avoids adverse selection while offering an effective insurance product involving a risk transfer for all policyholders. In a regime of algorithmic accountability, insurers should be held responsible for their pricing models in terms of transparency, fairness and solidarity. It is therefore very important to be able to ``look under the hood'' of machine learning algorithms and the resulting pricing models. That is exactly one of the goals of this paper.

In this paper, we study how tree-based machine learning methods can be applied to insurance pricing. The building blocks of these techniques are decision trees, covered in \citet{FriedmanESL}. These are simple predictive models that mimic human decision-making in the form of yes-no questions. In insurance pricing, a decision tree partitions a portfolio of policyholders into groups of homogeneous risk profiles based on some characteristics. The partition of the portfolio is directly observable, resulting in high transparency. For each subgroup, a constant prediction is put forward, automatically inducing solidarity among the policyholders in a subgroup (as long as the size of this subgroup is large enough). These aspects of decision trees make them good candidates for insurance pricing. However, the predictive performance of such simple trees tends to be rather low. We therefore consider more complex algorithms that combine multiple decision trees in an ensemble, i.e.,~tree-based machine learning. These ensemble techniques usually provide better predictive performance, but at the cost of less transparency. We employ model interpretation tools to understand these ``black boxes'', allowing us to interpret the resulting models and underlying decision process. Tree-based machine learning techniques are often praised for their ability to discover interaction effects in the data, a very useful insight for insurers that will be explored in this paper.

Insurance claim data typically entails highly imbalanced count data with excess zeros and varying exposure-to-risk on the frequency side, combined with long- or even heavy-tailed continuous data on the severity side. Standard machine learning algorithms typically deal with data that is more normal-like or balanced. \citet{Guelman2012} models the accident loss cost by simplifying the frequency count regression problem into a binary classification task. This however cannot factor in varying exposure-to-risk and leads to a loss of information regarding policyholders who file more than one claim in a period. \citet{Wuthrich2019} and \citet{Zochbauer2017} show how the specific data features on the frequency side can be taken into account for regression. 

We extend the existing literature by also putting focus on the severity side of claims and obtaining full tariff plans on real-world claims data from an insurance portfolio. We develop an elaborate cross-validation scheme instead of relying on built-in routines from software packages and we take into account multiple types of risk factors: categorical, continuous and spatial information. The goal of this paper is to investigate how tree-based pricing models perform compared to the classical actuarial approach with GLMs. This comparison puts focus on statistical performance, interpretation and business implications. We go beyond a purely statistical comparison, but acknowledge the fact that the resulting pricing model has to be deployed, after marketing adjustments, in a business environment with specific requirements.

The rest of this paper is structured as follows. Section~\ref{classicpriceframe} introduces the basic principles and guidelines for building a benchmark pricing GLM. Section~\ref{TBML} consolidates the important technical details on tree-based machine learning. Section~\ref{ResultFreqSev} presents interpretations from the optimal frequency and severity models fitted on a Belgian insurance data set, together with an out-of-sample performance comparison. Section~\ref{business} reviews the added value from a business angle and Section~\ref{conclusion} concludes this paper. In an accompanying online supplement, available at \url{https://github.com/henckr/sevtree}, we provide more details on the construction and interpretation of tree-based machine learning methods for the severity.

\section{State-of-the-art insurance pricing models}
\label{classicpriceframe}
To assess the possible merits of tree-based machine learning for insurance pricing, we first have to establish a fair benchmark pricing model that meets industry standards. GLMs are by far the most popular pricing models in today's industry. This section outlines the basic principles and steps for creating a benchmark pricing GLM with the strategy from \cite{Henckaerts2018}.

A P\&C insurance company is interested in the total loss amount $L$ per unit of exposure-to-risk~$e$, where $L$ is the total loss for the $N$ claims reported by a policyholder during the exposure period $e$. 
P\&C insurers usually opt for a so-called frequency-severity strategy to price a contract \citep{Denuit2007,Frees2014,Parodi2014}. Claim frequency $F$ is the number of claims $N$ filed per unit of exposure-to-risk $e$. Claim severity $S$ refers to the cost per claim and is defined by the average amount per claim filed, that is the total loss amount $L$ divided by the number of claims $N$. The technical price $\pi$ (or: pure/risk premium) then follows as:
\begin{equation*}
\pi \, = \, \mathbb{E}\left(\frac{L}{e}\right) \, \overset{\mathrm{indep.}}{=} \, \mathbb{E}\left(\frac{N}{e}\right) \times \mathbb{E}\left(\frac{L}{N} \mid N > 0\right) \, = \, \mathbb{E}(F) \times \mathbb{E}(S),
\label{tech_price}
\end{equation*}
assuming independence between the frequency and the severity component of the premium \citep{Klugman2012}. Alternatives, allowing dependence between $F$ and $S$, are investigated in the literature \citep{Gschlossl2007,Czado2012,Garrido2016}.

Predictive models for both $F$ and $S$ are typically developed within the framework of GLMs. Let $Y$, the response variable of interest, follow a distribution from the exponential family. The structure of a GLM with all explanatory variables in a categorical format is:
\begin{equation}
\eta = g(\mu) = \boldsymbol{z}^\top\boldsymbol{\beta} = \beta_0 + \sum_{j=1}^{q} \beta_j z_{j}\, ,
\label{linpredGLM}
\end{equation}
with $\eta$ the linear predictor, $g(\cdot)$ the link function and $\mu$ the expectation of $Y$. The $q+1$ dimensional 0/1-valued vector $\boldsymbol{z}$ contains a 1 for the intercept together with the $q$ dummy variables expressing the policyholder's risk profile. In a claim frequency model, the response variable $N$ typically follows a count distribution such as the Poisson. Assuming $g(\cdot) = \ln(\cdot)$, the model may account for exposure-to-risk through an offset $\ln(e)$ such that the risk premium is proportional to the exposure. In a claim severity model, the response variable $L/N$ typically follows a right skewed distribution with a long right tail such as the gamma or log-normal. Only policyholders filing at least one claim, i.e.,~$N>0$, contribute to the severity model calibration and the number of claims $N$ is used as a case weight in the regression \citep{Denuit2004}.

\begin{figure}[h!]
	\centering
	\begin{subfigure}{.333\textwidth}
		\renewcommand{\thesubfigure}{1a}
		\centering
		\includegraphics[width=0.86\textwidth]{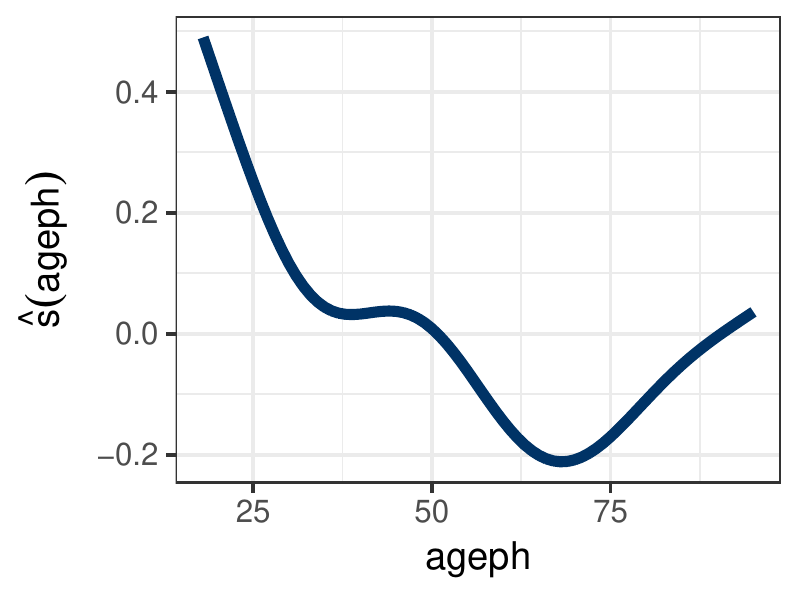}
		\caption{Smooth continuous effect}
		\label{bin1a}
	\end{subfigure}%
	\begin{subfigure}{.333\textwidth}
		\renewcommand{\thesubfigure}{1b}
		\centering
		\includegraphics[width=0.86\textwidth]{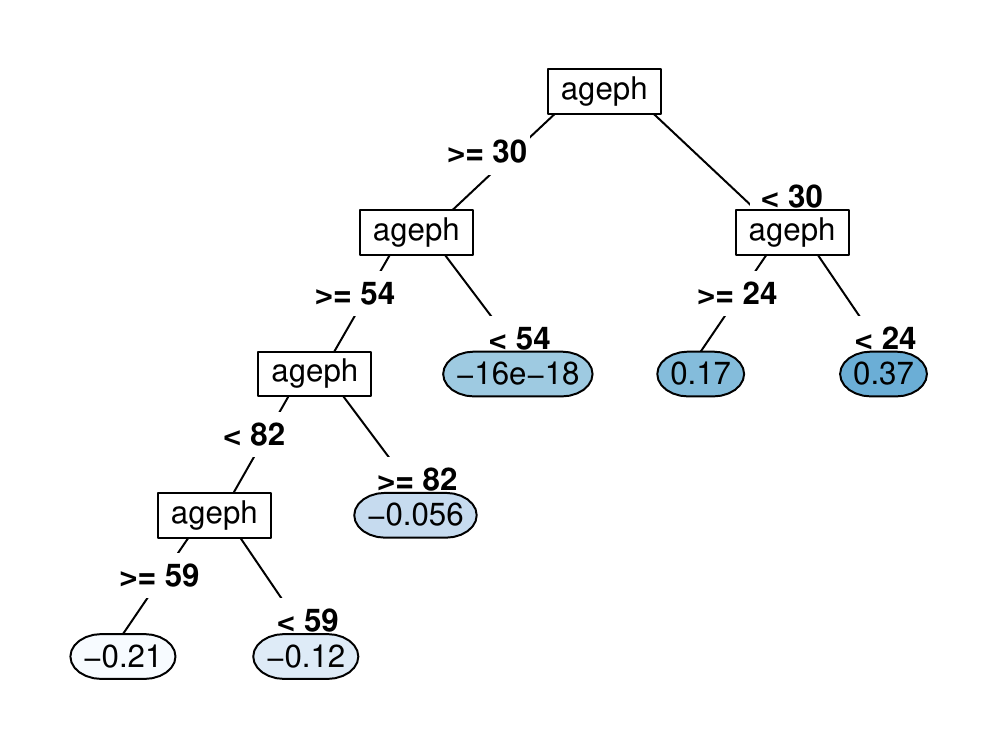}
		\caption{Supervised decision tree}
		\label{bin1b}
	\end{subfigure}%
	\begin{subfigure}{.333\textwidth}
		\renewcommand{\thesubfigure}{1c}
		\centering
		\includegraphics[width=0.86\textwidth]{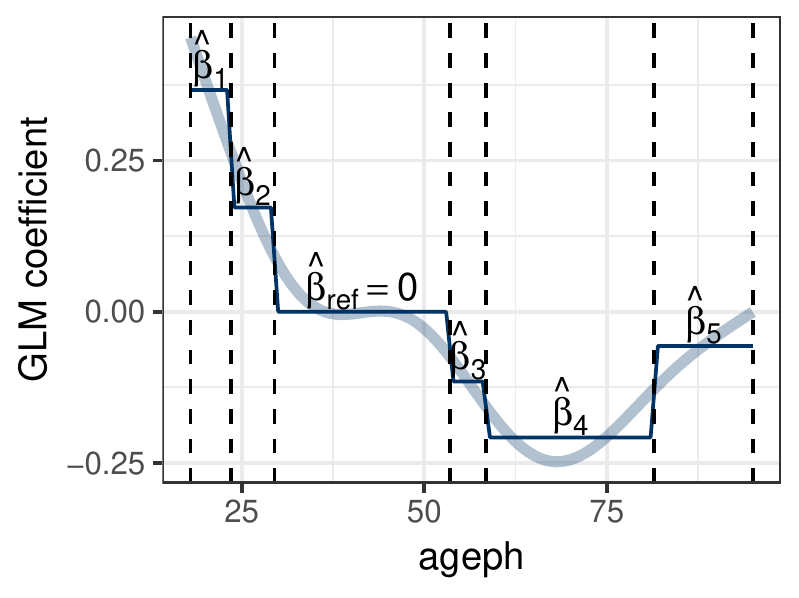}
		\caption{Binned continuous effect}
		\label{bin1c}
	\end{subfigure}
	
	\begin{subfigure}{.333\textwidth}
		\renewcommand{\thesubfigure}{2a}
		\centering
		\includegraphics[width=0.86\textwidth]{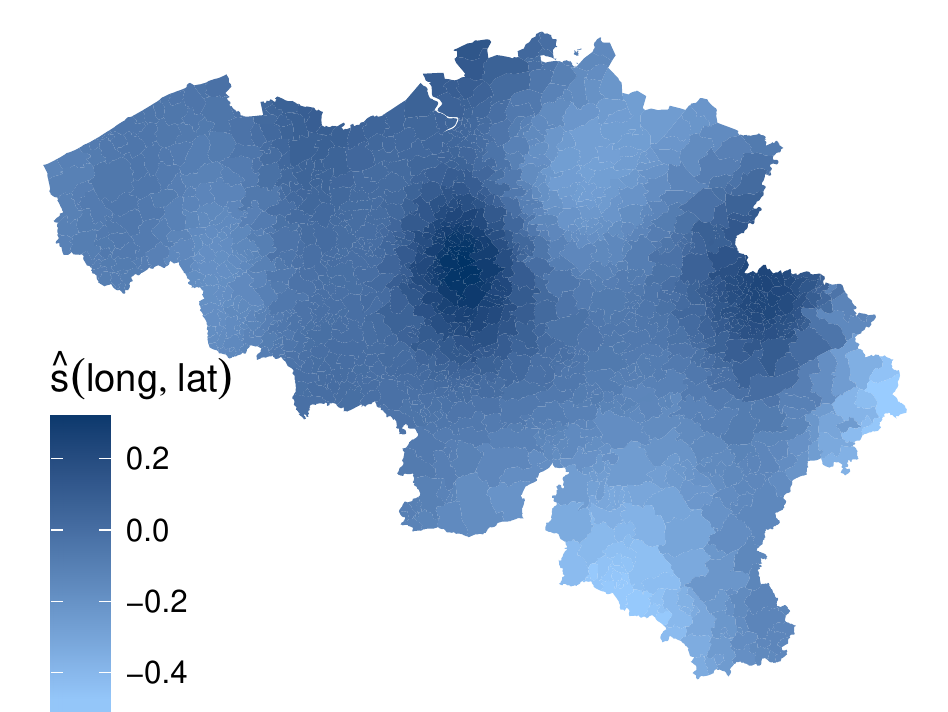}
		\caption{Smooth spatial effect}
		\label{bin2a}
	\end{subfigure}%
	\begin{subfigure}{.333\textwidth}
		\renewcommand{\thesubfigure}{2b}
		\centering
		\includegraphics[width=0.86\textwidth]{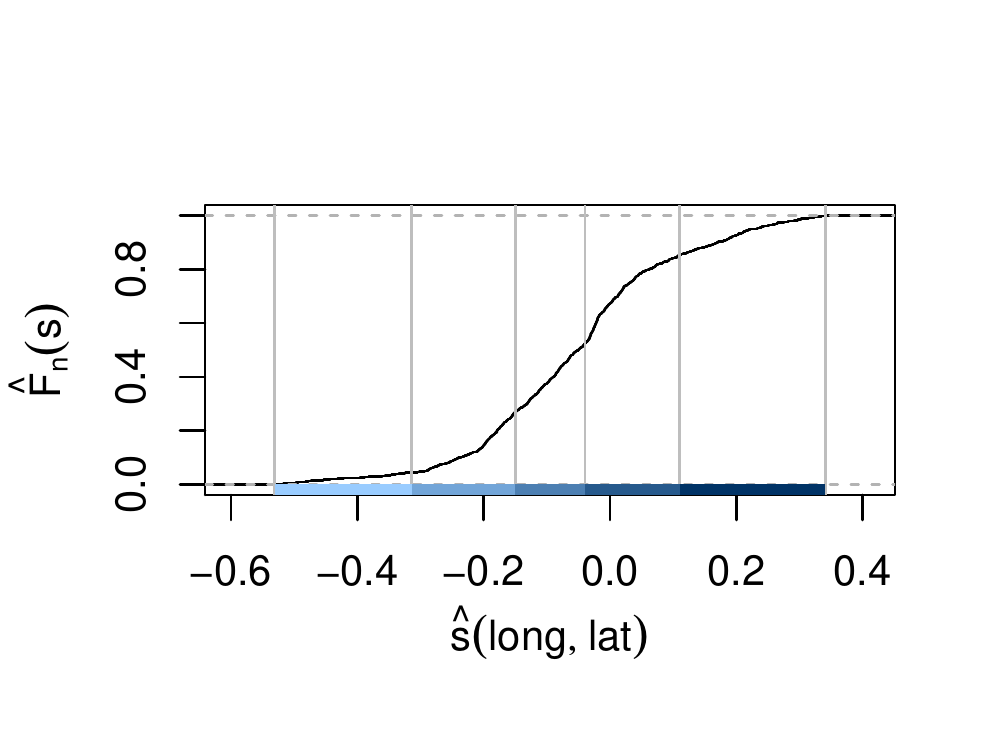}
		\caption{Unsupervised clustering}
		\label{bin2b}
	\end{subfigure}%
	\begin{subfigure}{.333\textwidth}
		\renewcommand{\thesubfigure}{2c}
		\centering
		\includegraphics[width=0.86\textwidth]{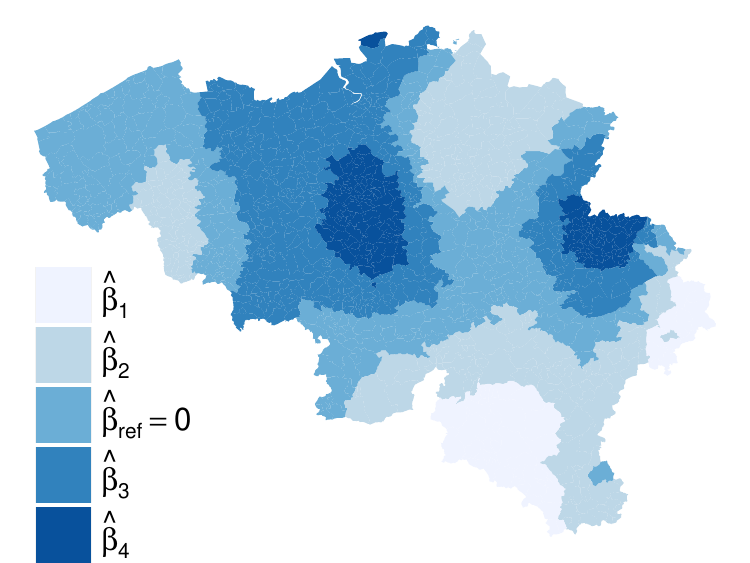}
		\caption{Binned spatial effect}
		\label{bin2c}
	\end{subfigure}%
	\caption{Schematic overview of the binning strategy of \citet{Henckaerts2018} for a continuous risk factor (\subref{bin1a}\,-\,\subref{bin1c}) and a spatial risk factor (\subref{bin2a}\,-\,\subref{bin2c}). The smooth GAM effect of a continuous risk factor (\subref{bin1a}) is fed as a response to a decision tree (\subref{bin1b}), which splits the continuous risk factor into bins (\subref{bin1c}). The smooth spatial effect (\subref{bin2a}) is clustered in an unsupervised way (\subref{bin2b}), resulting in groups of postcode areas (\subref{bin2c}). These categorical risk factors are used in a GLM.}
	\label{bin_schema} 
\end{figure}

\citet{Henckaerts2018} details a data-driven strategy to build a GLM with all risk factors in a categorical format. This strategy aligns the practical requirements of a business environment with the statistical flexibility of generalized additive models \citep[GAMs, documented by][]{Wood2006}. GAMs extend the linear predictor in Eq.~\eqref{linpredGLM} with (multidimensional) smooth functions. After an exhaustive search, the starting point for both frequency and severity is a flexible GAM with smooth effects for continuous risk factors, including two-way interactions, and a smooth spatial effect. These smooth effects are used to bin the continuous and spatial risk factors, thereby transforming them to categorical variables. Figure~\ref{bin_schema} schematizes how decision trees and unsupervised clustering are applied to achieve this binning. The output of this framework is an interpretable pricing GLM, which serves as benchmark pricing model in this study.

\section{Tree-based machine learning methods for insurance pricing}
\label{TBML}
Section~\ref{AlgoEssentials} introduces the essential algorithmic details needed for understanding the tree-based modeling techniques used in this paper. We consider regression trees \citep{Breiman1984}, random forests \citep{Breiman2001} and gradient boosting machines \citep{Friedman2001} as alternative modeling techniques for insurance pricing. These models rely on the choice of a loss function, which has to be tailored to the characteristics of insurance data as we motivate in Section~\ref{loss_fun}. In Section~\ref{Tuning} and \ref{interpret}, we explain our tuning strategy and present interpretation tools.

\subsection{Algorithmic essentials}
\label{AlgoEssentials}

\paragraph{Regression tree} Decision trees partition data based on yes-no questions, predicting the same value for each member of the constructed subsets. A popular approach to construct decision trees is the Classification And Regression Tree (CART) algorithm, introduced by \citet{Breiman1984}. The predictor space $R$ is the set of possible values for the $p$ variables $x_1,\ldots,x_p$, e.g.,~$R = \mathbb{R}^p$ for $p$ unbounded, continuous variables and $R = [\min(x_1),\max(x_1)] \times [\min(x_2),\max(x_2)]$ for two bounded, continuous variables. A tree divides the predictor space $R$ into $J$ distinct, non-overlapping regions $R_1,\ldots,R_J$. In the $j$th region, the fitted response $\hat{y}_{R_j}$ is computed as a (weighted) average of the training observations falling in that region. The regression tree predicts a (new) observation with characteristics $\boldsymbol{x}$ as follows:
\begin{equation}
f_{\text{tree}}(\boldsymbol{x}) = \sum_{j=1}^{J} \hat{y}_{R_j} \, \mathbbm{1}(\boldsymbol{x}\in R_j).
\label{pred_tree}
\end{equation}
The indicator $\mathbbm{1}(A)$ equals one if event $A$ occurs and zero otherwise.
As the $J$ regions are non-overlapping, the indicator function differs from zero for exactly one region for each $\boldsymbol{x}$. A tree therefore makes the same constant prediction $\hat{y}_{R_j}$ for the entire region $R_j$.

It is computationally impractical to consider every possible partition of the predictor space~$R$ in $J$ regions, therefore CART uses a top-down greedy approach known as recursive binary splitting. From the full predictor space $R$, the algorithm selects a splitting variable $x_v$ with $v \in \{1, \ldots, p\}$ and a cut-off $c$ such that $R = R_1(v,c) \cup R_2(v,c)$ with $R_1(v,c) = \left\lbrace R \, | \, x_v \leqslant c \right\rbrace$ and $R_2(v,c) = \left\lbrace R \, | \, x_v > c \right\rbrace$. This forms two nodes in the tree, one containing the observations satisfying $x_v \leqslant c$ and the other containing the observations satisfying $x_v > c$. For a categorical splitting variable, the corresponding factor levels are replaced by their empirical response averages, see Section~8.8 in \citet{Breiman1984}. These averages are sorted from low to high and a cut-off $c$ is chosen such that the factor levels are split into two groups. The splitting variable $x_v$ and cut-off $c$ are chosen such that their combination results in the largest improvement in a carefully picked loss function $\mathscr{L}(\cdot\, ,\cdot)$. For $i=\{1,\ldots,n\}$, where $n$ is the number of observations in the training set, the CART algorithm searches for $x_v$ and $c$ minimizing the following summations:
\begin{equation*}
\underset{i\,:\,\boldsymbol{x}_i \in R_1(v,c)}{\sum} \mathscr{L}(y_i,\hat{y}_{R_1})  \,\,\, + \underset{i\,:\,\boldsymbol{x}_i \in R_2(v,c)}{\sum} \mathscr{L}(y_i,\hat{y}_{R_2}).
\end{equation*}
A standard loss function is the squared error loss, but we present more suitable loss functions for claim frequency or severity data in Section~\ref{loss_fun}. In a next iteration, the algorithm splits $R_1$ and/or $R_2$ in two regions and this process is repeated recursively until a stopping criterion is satisfied. This stopping criterion typically puts a predefined limit on the size of a tree, e.g.,~a minimum improvement in the loss function \citep{Breiman1984}, a maximum depth of the tree or a minimum number of observations in a node of the tree \citep{FriedmanESL}.

A large tree is likely to overfit the data and does not generalize well to new data, while a small tree is likely to underfit the data and fails to capture the general trends. This is related to the bias-variance tradeoff \citep{FriedmanESL} meaning that a large tree has low bias and high variance while a small tree has high bias but low variance. To prevent overfitting, the performance of a tree is penalized by the number of regions $J$ as follows:
\begin{equation}
\sum_{j=1}^{J} \, \underset{i\,:\,\boldsymbol{x}_i \in R_j}{\sum}\mathscr{L}(y_i,\hat{y}_{R_j}) \,\,\, + \,\,\,  J \cdot cp \cdot \underset{i\,:\,\boldsymbol{x}_i \in R}{\sum}\mathscr{L}(y_i,\hat{y}_{R})\,,
\label{tree_cp}
\end{equation}
where the first part assesses the goodness of fit and the second part is a penalty measuring the tree complexity. The strength of this penalty is driven by the complexity parameter $cp$, a tuning parameter (see Section~\ref{Tuning} for details on the tuning strategy). A large (small) value for $cp$ puts a high (low) penalty on extra splits and will result in a small (large) tree. The complexity parameter $cp$ is usually scaled with the loss function evaluated for the root tree, which is exactly the last summation in Eq.~\eqref{tree_cp}; see Remark 3.8 in \citet{Zochbauer2017}. This ensures that $cp = 1$ delivers a root tree without splits capturing an overall $y$ estimate (denoted $\hat{y}_R$ in Eq.~\eqref{tree_cp}) and $cp = 0$ results in the largest possible tree allowed by the stopping criterion.

\begin{figure}[h!]
	\centering
	\includegraphics[width=0.94\textwidth]{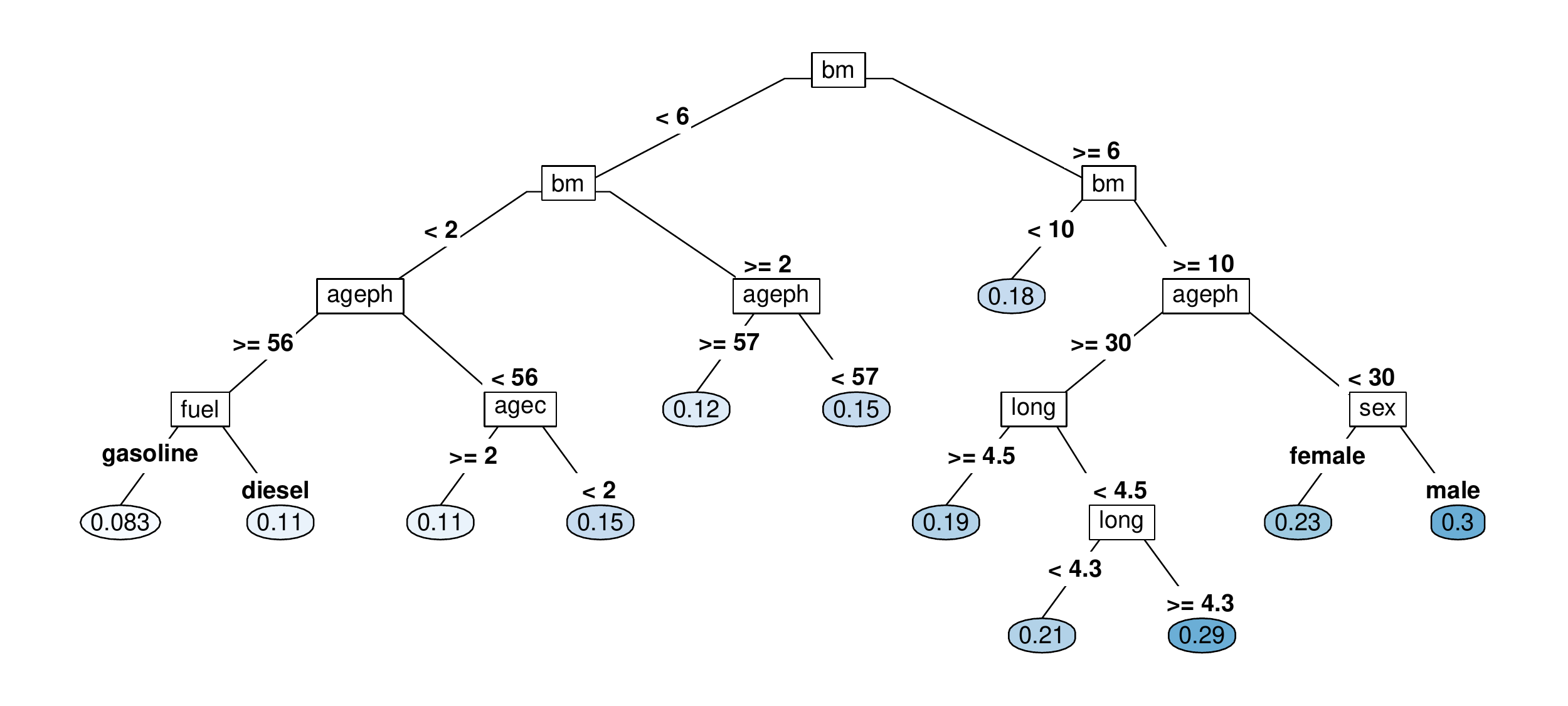}
	\caption{Visual representation of a regression tree for claim frequency with nodes (rectangles) containing the splitting variable $x_v$, edges (lines) representing the splits with cut-off $c$ and leaf nodes (ellipses) containing the prediction values $\hat{y}_{R_j}$. The variable names are defined in Table~\ref{variables}.}
	\label{reg_tree}
\end{figure}

Figure~\ref{reg_tree} depicts an example of a regression tree for claim frequency data. The rectangles are internal nodes which partition observations going from top to bottom along the tree. The top node, splitting on the bonus-malus level \texttt{bm}, is called the root node. The ellipses are leaf nodes, containing prediction values for the observations ending in that specific leaf. Going from left to right, the leaf nodes are ordered from low (light blue) to high (dark blue) prediction values. Decision trees have many advantages because their predictions are highly explainable and interpretable, both very important criteria for regulators. The downside of trees however is that the level of predictive accuracy tends to be lower compared to other modeling techniques. This is mainly driven by the high variance of a tree, e.g.,~slight changes in the data can result in very different trees and therefore rather different predictions for certain observations. The predictive performance can be substantially improved by aggregating multiple decision trees in ensembles of trees, thereby reducing the variance. That is the idea behind popular ensemble methods, such as random forests and gradient boosting machines, which are discussed next.

\paragraph{Random forest} Bagging, which is short for \textbf{b}ootstrap \textbf{agg}regat\textbf{ing} \citep{Breiman1996}, and random forests \citep{Breiman2001} are similar ensemble techniques combining multiple decision trees. Bagging reduces the variance of a single tree by averaging the forecasts of multiple trees on bootstrapped samples of the original data. This stabilizes the prediction and improves the predictive performance compared to a single decision tree. Starting from the data set $\mathcal{D}$, the idea of bagging is to take bootstrap samples $\{\mathcal{D}_t\}_{t=1,\ldots,T}$ and to build $T$ decision trees, one for each $\mathcal{D}_t$ independently. The results are then aggregated in the following way:
\begin{equation}
f_{\text{bagg}}(\boldsymbol{x}) = \frac{1}{T} \sum_{t=1}^{T} f_{\text{tree}}(\boldsymbol{x} \, | \, \mathcal{D}_t)\,,
\label{pred_bagg}
\end{equation}
where the condition $(| \, \mathcal{D}_t)$ indicates that the tree was developed on the sample $\mathcal{D}_t$. 

The performance improvement through variance reduction gets bigger when there is less correlation between the individual trees, see Lemma 3.25 in \cite{Zochbauer2017}. For that reason, the trees are typically grown deep (i.e., $cp = 0$ in Eq.~\eqref{tree_cp}), until a stopping criterion is satisfied. Taking bootstrap samples of smaller sizes $\delta \cdot n$, with $n$ the number of observations in $\mathcal{D}$ and $0 < \delta < 1$, decorrelates the trees further and reduces the model training time. However, a lot of variability remains within a bagged ensemble because the trees built on the bootstrapped data samples are still quite similar. This is especially the case when some explanatory variables in the data are much more predictive than the others. The important variables will dominate the first splits, causing all trees to be similar to one another. To prevent this, a random forest further decorrelates the individual trees by sampling variables during the growing process. At each split, $m$ out of $p$ variables are randomly chosen as candidates for the optimal splitting variable. Besides this adaptation, a random forest follows the same strategy as bagging and predicts a new observation according to Eq.~\eqref{pred_bagg}. The random forest procedure is detailed in Algorithm \ref{rf_algo} where $T$ and~$m$ are treated as tuning parameters (see Section~\ref{Tuning} for details on the tuning strategy).

\begin{algorithm}[H]
	\vspace{-0mm} \hrulefill \\
	\For{$t=1,\ldots,T$}{
		generate bootstrapped data $\mathcal{D}_t$ of size $\delta \cdot n$ by sampling with replacement from data $\mathcal{D}$\;
		\While{stopping criterion not satisfied}{
			randomly select $m$ of the $p$ variables\;
			find the optimal splitting variable $x_v$ from the $m$ options together with cut-off $c$\;
		}
	}
	$f_{\text{rf}}(\boldsymbol{x}) = \frac{1}{T} \sum_{t=1}^{T} f_{\text{tree}}(\boldsymbol{x} \, | \, \mathcal{D}_t)$\;
	\vspace{-3mm} \hrulefill \vspace{2mm}
	\caption{Procedure to build a random forest model.}
	\label{rf_algo}
\end{algorithm}

A random forest improves the predictive accuracy obtained with a single decision tree by using more, and hopefully slightly different, trees to solve the problem at hand. However, the trees in a random forest are built independently from each other (i.e.,~ the \texttt{for} loop in Algorithm \ref{rf_algo} can be run in parallel) and do not share information during the training process. 

\paragraph{Gradient boosting machine} In contrast to random forests, boosting is an iterative statistical method that combines many weak learners into one powerful predictor. \citet{Friedman2001} introduced decision trees as weak learners; each tree improves the current model fit, thereby using information from previously grown trees. At each iteration, the pseudo-residuals are used to assess the regions of the predictor space for which the model performs poorly in order to improve the fit in a direction of better overall performance. The pseudo-residual $\rho_{i,t}$ for observation $i$ in iteration $t$ is calculated as the negative gradient of the loss function~$-\partial \mathscr{L}\{y_i,f(\boldsymbol{x}_i)\} / \partial f(\boldsymbol{x}_i)$, evaluated at the current model fit. This typical approach called stepwise gradient descent ensures that a lower loss is obtained at the next iteration, until convergence. The boosting method learns slowly by fitting a small tree of depth $d$ (with a squared error loss function) to these pseudo-residuals, improving the model fit in areas where it does not perform well. For each region $R_j$ of that tree, the update $\hat{b}_j$ is calculated as the constant that has to be added to the previous model fit to minimize the loss function, namely $b$ that minimizes $\mathscr{L}\{y_i,f(\boldsymbol{x}_i)+b\}$ over this region. A shrinkage parameter $\lambda$ controls the learning speed by shrinking updates for $\boldsymbol{x} \in R_j$ as follows: $f_{new}(\boldsymbol{x}) = f_{old}(\boldsymbol{x}) + \lambda \cdot \hat{b}_j$. A lower $\lambda$ usually results in better performance but also increases computation time because more trees are needed to converge to a good solution. Typically, $\lambda$ is fixed at the lowest possible value within the computational constraints \citep{Friedman2001}. The collection of $T$ trees at the final iteration is used to make predictions.

Stochastic gradient boosting, introduced by \citet{Friedman2002}, injects randomness in the training process. In each iteration, the model update is computed from a randomly selected subsample of size $\delta \cdot n$. This improves both the predictive accuracy and model training time when $\delta < 1$. 
The (stochastic) gradient boosting machine algorithm is given in Algorithm \ref{gbm_algo} where $T$ and $d$ are treated as tuning parameters (see Section~\ref{Tuning} for details on the tuning strategy).

\begin{algorithm}[H]
	\vspace{-0mm} \hrulefill \\
	initialize fit to the optimal constant model: $f_0(\boldsymbol{x}) = \arg \min_b \sum_{i=1}^{n} \mathscr{L}(y_i,b)$\;
	\For{$t=1,\ldots, T$}{
		randomly subsample data of size $\delta \cdot n$ without replacement from data $\mathcal{D}$\;
		\For{$i=1,\ldots,\delta \cdot n$}{\vspace{3mm} $\rho_{i,t} = -\left[ \frac{\partial \mathscr{L}\{y_i,f(\boldsymbol{x}_i)\}}{\partial f(\boldsymbol{x}_i)} \right]_{f=f_{t-1}}$}
		fit a tree of depth $d$ to the pseudo-residuals $\rho_{i,t}$ resulting in regions $R_{j,t}$ for $j=1,\ldots,J_t$\;
		\For{$j=1,\ldots,J_t$}{ \vspace{3mm} $\hat{b}_{j,t} = \arg \min_b \underset{i\,:\,\boldsymbol{x}_i \in R_{j,t}}{\sum} \mathscr{L}\{y_i,f_{t-1}(\boldsymbol{x}_i) + b\}$}
		update $f_t(\boldsymbol{x}) = f_{t-1}(\boldsymbol{x}) + \lambda \sum_{j=1}^{J_t} \hat{b}_{j,t}  \mathbbm{1}(\boldsymbol{x} \,\in R_{j,t})$\;
	}
	$f_{\text{gbm}}(\boldsymbol{x}) = f_T(\boldsymbol{x})$\;
	\vspace{-3mm} \hrulefill \vspace{2mm}
	\caption{Procedure to build a (stochastic) gradient boosting machine.}
	\label{gbm_algo}
\end{algorithm}

\subsection{Loss functions for insurance data}
\label{loss_fun}
The machine learning algorithms discussed in Section~\ref{AlgoEssentials} require the specification of a loss (or: cost) function that is to be minimized during the training phase of the model. We first present a general discussion on the loss function choice, followed by details on the \textsf{R} implementation.

\paragraph{Loss functions} The standard loss function for regression problems is the squared error loss:
\begin{equation*}
\mathscr{L}\{y_i,f(\boldsymbol{x_i})\} \propto \{y_i - f(\boldsymbol{x}_i)\}^2\, ,
\end{equation*}
where $y_i$ is the observed response and $f(\boldsymbol{x}_i)$ is the prediction of the model for variables $\boldsymbol{x}_i$. However, the squared error is not necessarily a good choice when modeling integer-valued frequency data or right-skewed severity data. We use the concept of deviance to make this idea clear. The deviance is defined as $D\{y,f(\boldsymbol{x})\} = -2 \cdot \ln[\mathcal{L}\{f(\boldsymbol{x})\}/ \mathcal{L}(y)]$, a likelihood ratio where $\mathcal{L}\{f(\boldsymbol{x})\}$ is the model likelihood and $\mathcal{L}(y)$ the likelihood of the saturated model (i.e.,~the model in which the number of parameters equals the number of observations). The condition $\mathcal{L}\{f(\boldsymbol{x})\} \leqslant \mathcal{L}(y)$ always holds, so the ratio of likelihoods is bounded from above by one. For competing model fits, the best one obtains the lowest deviance value on holdout data. We therefore use a loss function $\mathscr{L}(\cdot\, , \cdot)$ such that $D\{y,f(\boldsymbol{x})\} = \sum_{i=1}^{n} \mathscr{L}\{y_i,f(\boldsymbol{x}_i)\}$. This idea was put forward by \citet{Venables2002} for general classification and regression problems.

Assuming constant variance, the normal (or: Gaussian) deviance can be expressed as follows:
\begin{align*}
D\{y,f(\boldsymbol{x})\} & = 2 \ln \prod_{i=1}^{n} \exp \left\{- \frac{1}{2\sigma^2}(y_i-y_i)^2\right\} - 2 \ln \prod_{i=1}^{n} \exp \left[- \frac{1}{2\sigma^2}\{y_i-f(\boldsymbol{x}_i)\}^2\right] \notag \\
& = \frac{1}{\sigma^2} \sum_{i=1}^{n}\{y_i - f(\boldsymbol{x}_i)\}^2 \, ,
\end{align*}
which boils down to a scaled version of the sum of squared errors. This implies that a loss function based on the squared error is appropriate when the normal assumption is reasonable. More generally, the squared error is suitable for any continuous distribution symmetrical around its mean with constant variance, i.e.,~any elliptical distribution. However, claim frequency and severity data do not follow any elliptical distribution, as we show in Section~\ref{data}. Therefore, in an actuarial context, \citet{Wuthrich2019} and \citet{Zochbauer2017} propose more suitable loss functions inspired by the GLM pricing framework from Section~\ref{classicpriceframe}.

Claim frequency modeling involves count data, typically assumed to be Poisson distributed in GLMs. Therefore, an appropriate loss function is the Poisson deviance, defined as follows:
\begin{align}
D(y,f(\boldsymbol{x})) & = 2 \ln \prod_{i=1}^{n} \exp(-y_i)\frac{y_i^{y_i}}{y_i!} - 2 \ln \prod_{i=1}^{n} \exp\{-f(\boldsymbol{x}_i)\}\frac{f(\boldsymbol{x}_i)^{y_i}}{y_i!} \notag \\
& = 2 \sum_{i=1}^{n} \left[ y_i \ln \frac{y_i}{f(\boldsymbol{x}_i)} - \{y_i - f(\boldsymbol{x}_i)\} \right].
\label{poiss_dev}
\end{align}
When using an exposure-to-risk measure $e_i$, $f(\boldsymbol{x}_i)$ is replaced by $e_i \cdot f(\boldsymbol{x}_i)$ such that the exposure is taken into account in the expected number of claims. Thus, the Poisson deviance loss function can account for different policy durations. Predictions from a Poisson regression tree in Eq.~\eqref{pred_tree} are equal to the sum of the number of claims divided by the sum of exposure for all training observations in each leaf node: $\hat{y}_{R_j} = \sum_{i\in I_j}N_i / \sum_{i\in I_j} e_i$ for $I_j = \{i: \boldsymbol{x}_i\in R_j\}$. This optimal estimate is obtained by setting the derivative of Eq.~\eqref{poiss_dev} with respect to $f$ equal to zero. As a tree node without claims leads to a division by zero in the deviance calculation, an adjustment can be made to the implementation with a hyper-parameter that will be introduced in Section~\ref{Tuning}.

Right-skewed and long-tailed severity data is typically assumed to be gamma or log-normally distributed in GLMs. In Section~\ref{ResultFreqSev}, we present the results obtained with the gamma deviance as our preferred model choice, but a discussion on the use of the log-normal deviance is available in the supplementary material. The gamma deviance is defined as follows:
\begin{align}
D\{y,f(\boldsymbol{x})\} & = 2 \ln \prod_{i=1}^{n} \frac{1}{y_i \Gamma(\alpha)} \left(\frac{\alpha y_i}{y_i}\right)^\alpha \exp\left(-\frac{\alpha y_i}{y_i}\right) - 2 \ln \prod_{i=1}^{n} \frac{1}{y_i \Gamma(\alpha)} \left\{\frac{\alpha y_i}{ f(\boldsymbol{x}_i)}\right\}^\alpha \exp\left\{-\frac{\alpha y_i}{ f(\boldsymbol{x}_i)}\right\} \notag \\
& = 2 \sum_{i=1}^{n} \alpha \left\{ \frac{y_i - f(\boldsymbol{x}_i)}{f(\boldsymbol{x}_i)} - \ln \frac{y_i}{f(\boldsymbol{x}_i)} \right\}.
\label{gamma_dev}
\end{align}
The shape parameter $\alpha$ acts as a scaling factor and can therefore be ignored. When dealing with case weights, $\alpha$ can be replaced by the weights $w_i$. In severity modeling, the response is the average claim amount of the $N_i$ observed claims and the number of claims $N_i$ should be used as case weight. Predictions from a gamma regression tree in Eq.~\eqref{pred_tree} are equal to the sum of the total loss amount divided by the sum of the number of claims for all training observations in each leaf node: $\hat{y}_{R_j} = \sum_{i\in I_j}L_i / \sum_{i\in I_j} N_i$ for $I_j = \{i: \boldsymbol{x}_i\in R_j\}$. This optimal estimate is obtained by setting the derivative of Eq.~\eqref{gamma_dev} with respect to $f$ equal to zero.

\paragraph{Implementation in \textsf{R}} Our results are obtained with two special purpose packages for tree-based machine learning in the statistical software \textsf{R}. For the regression trees and random forests, we developed our own package called \texttt{distRforest} \citep{Henckaerts2019}. For stochastic gradient boosting, we chose the implementation from \citet{Southworth2015} of the \texttt{gbm} package, originally developed by \citet{Ridgeway2014}. Our \texttt{distRforest} package extends the \texttt{rpart} package by \citet{Therneau2018} such that it is capable of developing regression trees and random forests with our specific desired loss functions for both claim frequency and severity. We had to go beyond the standard implementations especially because of the loss functions appropriate for actuarial applications. Although the \texttt{rpart} package supports the Poisson deviance for regression trees, it did not facilitate the use of a suitable loss function for severity data.

\subsection{Tuning strategy}
\label{Tuning}
\paragraph{Tuning and hyper-parameters} Table~\ref{params} gives an overview of the parameters used by the algorithms described in Section~\ref{AlgoEssentials}. Some of these are chosen with care (tuning parameters), while others are less influential and are set to a sensible predetermined value (hyper-parameters). Instead of relying on the built-in tuning strategies of the \textsf{R} packages mentioned in Section~\ref{loss_fun}, we perform an extensive grid search to find the optimal values among a predefined tuning grid displayed in Table~\ref{grid} in Appendix \ref{App_grid}. We prefer a grid search above other tuning strategies, such as Bayesian optimization \citep{Xia2017}, for its ease of implementation while being a sound approach. The hyper-parameter $\kappa$ enforces a stopping criterion for trees used across the three algorithms, ensuring that a split is not allowed if a resulting node would contain less than $1\%$ of the observations. The hyper-parameter $\delta$ in Algorithms~\ref{rf_algo} and \ref{gbm_algo} specifies to develop the trees in the ensemble techniques on $75\%$ of the available training data. The shrinkage parameter $\lambda$ in Algorithm~\ref{gbm_algo} is set at a low value for which computation time is still reasonable, namely \num{0.01}. The parameter $\gamma$ helps to avoid division by zero when optimizing the Poisson deviance in Eq.~\eqref{poiss_dev}. This parameter is therefore only used when growing a regression tree and random forest for claim frequency. We refer the reader to Section 8.2 in \citet{Therneau1997} for details on the \texttt{rpart} implementation. In short, a gamma prior is assumed on the Poisson rate parameter to keep it from becoming zero when there is no claim in a node. With $I_j = \{i: \boldsymbol{x}_i\in R_j\}$, the prediction in a node is adapted as follows:
	\begin{equation*}
	\hat{y}^\gamma_{R_j} = \frac{\gamma^{-2} + \sum_{i\in I_j}N_i}{\gamma^{-2} / \hat{y}_{R}  + \sum_{i\in I_j} e_i}  \, .
	\end{equation*}
Note that $\hat{y}^\gamma_{R_j} = \hat{y}_{R}$ for $\gamma = 0$ and $\hat{y}^\gamma_{R_j} = \hat{y}_{R_j} = \sum_{i\in I_j}N_i / \sum_{i\in I_j} e_i$ for $\gamma = \infty$.

\begin{table}[h!]
	\centering
	\begin{tabular}{lcr}
		\toprule
		& Tuning parameters & Hyper-parameters \\
		\midrule
		\multirow{2}{*}{Regression tree}  & complexity parameter $cp$  & $\kappa = 0.01$  \\
		& {coefficient of variation gamma prior $\gamma$}  & \\
		\midrule
		\multirow{2}{*}{Random forest}  & number of trees $T$  & $cp = 0$ \, $\gamma$ = 0.25 \\
		& number of split candidates $m$ & $\kappa = 0.01$ \, $\delta$ = 0.75  \\
		\midrule
		\multirow{2}{*}{Gradient boosting machine} & number of trees $T$  & $\lambda = 0.01$ \\
		& tree depth $d$ & $\kappa = 0.01$ \, $\delta$ = 0.75   \\
		\bottomrule
	\end{tabular}
	\caption{Overview of the tuning and hyper-parameters for the different machine learning techniques.}
	\label{params}
\end{table}

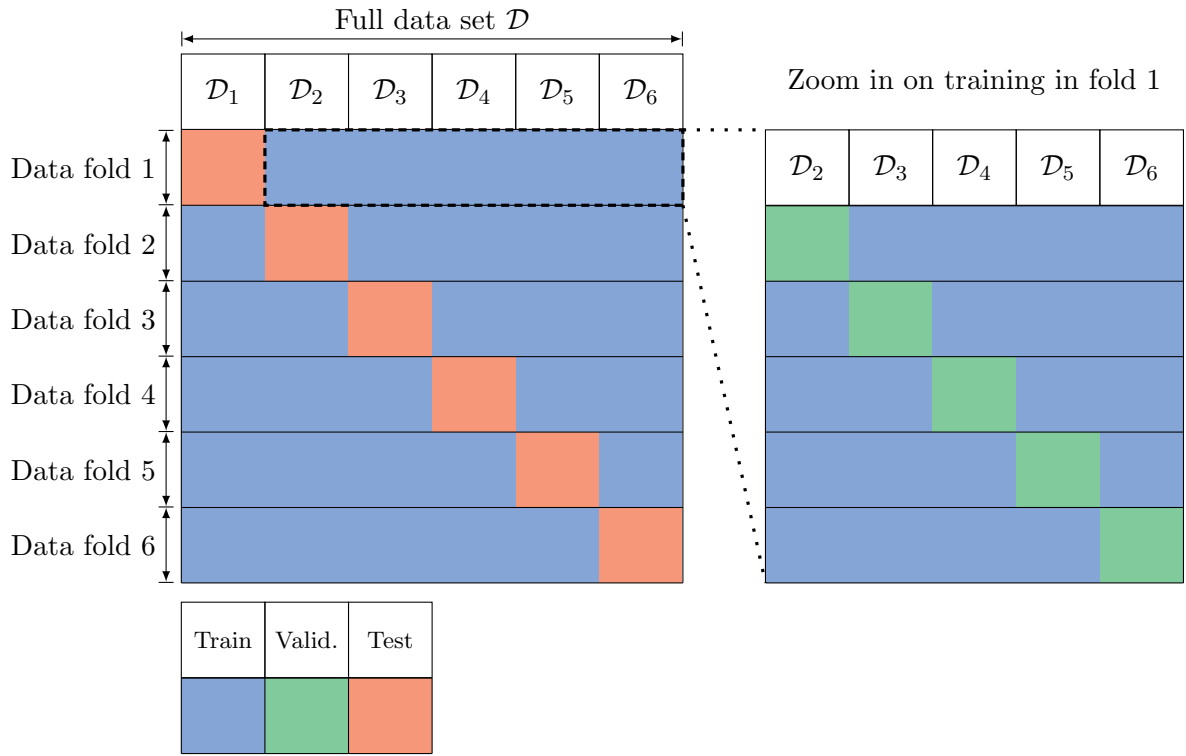
\begin{figure}[h!]
	\centering
	\begin{tikzpicture}
	\matrix (M) [matrix of nodes,
	nodes={minimum height = 1cm, minimum width = 1.1cm, outer sep=0, anchor=center, draw},
	column sep=-\pgflinewidth, 
	g/.style={fill=Green!50, draw=none},
	b/.style={fill=NavyBlue!50, draw=none},
	r/.style={fill=Red!50, draw=none},
	w/.style={fill=White!50, draw=none}
	]
	{
		{$\mathcal{D}_1$} & {$\mathcal{D}_2$} & {$\mathcal{D}_3$} & {$\mathcal{D}_4$} & {$\mathcal{D}_5$} & {$\mathcal{D}_6$} &  & |[w]|  & |[w]| & |[w]| & |[w]| & |[w]| \\[-0.05ex]
		|[r]| & |[b]| & |[b]| & |[b]| & |[b]| & |[b]| & |[w]| & {$\mathcal{D}_2$} & {$\mathcal{D}_3$} & {$\mathcal{D}_4$} & {$\mathcal{D}_5$} & {$\mathcal{D}_6$} \\
		|[b]| & |[r]| & |[b]| & |[b]| & |[b]| & |[b]| & |[w]| & |[g]| & |[b]| & |[b]| & |[b]| & |[b]| \\
		|[b]| & |[b]| & |[r]| & |[b]| & |[b]| & |[b]| & |[w]| & |[b]| & |[g]| & |[b]| & |[b]| & |[b]| \\
		|[b]| & |[b]| & |[b]| & |[r]| & |[b]| & |[b]| & |[w]| & |[b]| & |[b]| & |[g]| & |[b]| & |[b]| \\
		|[b]| & |[b]| & |[b]| & |[b]| & |[r]| & |[b]| & |[w]| & |[b]| & |[b]| & |[b]| & |[g]| & |[b]| \\
		|[b]| & |[b]| & |[b]| & |[b]| & |[b]| & |[r]| & |[w]| & |[b]| & |[b]| & |[b]| & |[b]| & |[g]|  \\[1.5ex]
		{\footnotesize Train}&{\footnotesize Valid.}&{\footnotesize Test}&&&&&&&&\\
		|[b]|&|[g]|&|[r]|&&&\\
	};
	
	\draw (M-1-1.north west) ++(0,2mm) coordinate (LT) edge[|<->|, >= latex] node[above]{Full data set $\mathcal{D}$} (LT-|M-1-6.north east);
	\draw[|<->|, >= latex, transform canvas={xshift=-2mm}] (M-2-1.north west) -- (M-2-1.south west) node[midway,left]{Data fold 1};
	\draw[<->|, >= latex, transform canvas={xshift=-2mm}] (M-3-1.north west) -- (M-3-1.south west) node[midway,left]{Data fold 2};
	\draw[<->|, >= latex, transform canvas={xshift=-2mm}] (M-4-1.north west) -- (M-4-1.south west) node[midway,left]{Data fold 3};
	\draw[<->|, >= latex, transform canvas={xshift=-2mm}] (M-5-1.north west) -- (M-5-1.south west) node[midway,left]{Data fold 4};
	\draw[<->|, >= latex, transform canvas={xshift=-2mm}] (M-6-1.north west) -- (M-6-1.south west) node[midway,left]{Data fold 5};
	\draw[<->|, >= latex, transform canvas={xshift=-2mm}] (M-7-1.north west) -- (M-7-1.south west) node[midway,left]{Data fold 6};
	
	\node[fit=(M-1-8)(M-1-12)]{Zoom in on training in fold 1};
	\draw[very thick, densely dashed, black] (M-2-2.north west) -- (M-2-6.north east) -- (M-2-6.south east) -- (M-2-2.south west) -- (M-2-2.north west)  ;
	\draw[loosely dotted, very thick] (M-2-6.north east) -- (M-2-8.north west);
	\draw[loosely dotted, very thick] (M-2-6.south east) -- (M-7-8.south west);
	
	\draw[] (M-2-1.north west) -- (M-7-1.south west);
	\draw[] (M-2-6.north east) -- (M-7-6.south east);
	\draw[] (M-7-1.south west) -- (M-7-6.south east);
	\draw[] (M-3-8.north west) -- (M-7-8.south west);
	\draw[] (M-3-12.north east) -- (M-7-12.south east);
	\draw[] (M-3-8.north west) -- (M-3-12.north east);
	\draw[] (M-7-8.south west) -- (M-7-12.south east);
	\draw[] (M-9-1.north west) -- (M-9-1.south west);
	\draw[] (M-9-3.north east) -- (M-9-3.south east);
	\draw[] (M-9-1.north west) -- (M-9-3.north east);
	\draw[] (M-9-1.south west) -- (M-9-3.south east);
	\draw[] (M-9-1.north east) -- (M-9-1.south east);
	\draw[] (M-9-2.north east) -- (M-9-2.south east);
	
	\draw[] (M-2-1.south west) -- (M-2-6.south east);
	\draw[] (M-3-1.south west) -- (M-3-6.south east);
	\draw[] (M-4-1.south west) -- (M-4-6.south east);
	\draw[] (M-5-1.south west) -- (M-5-6.south east);
	\draw[] (M-6-1.south west) -- (M-6-6.south east);
	\draw[] (M-3-8.south west) -- (M-3-12.south east);
	\draw[] (M-4-8.south west) -- (M-4-12.south east);
	\draw[] (M-5-8.south west) -- (M-5-12.south east);
	\draw[] (M-6-8.south west) -- (M-6-12.south east);
	
	\end{tikzpicture}
	\caption{Graphical representation of the cross-validation scheme. The holdout test set for data fold $k$ is $\mathcal{D}_k$, indicated in red. Within data fold $k$, we tune the parameters by 5-fold cross-validation on $\mathcal{D} \setminus \mathcal{D}_k$ with the validation sets $\mathcal{D}_\ell$ in green and the training data $\mathcal{D} \setminus \left\lbrace \mathcal{D}_k,\mathcal{D}_\ell \right\rbrace $ in blue. After tuning, we train the model on $\mathcal{D} \setminus \mathcal{D}_k$ using the optimal parameters for data fold $k$.}
	\label{cv_graph}
\end{figure}

\paragraph{Cross-validation} Machine learning typically relies on training data to build a model, validation data to tune the parameters and test data to evaluate the out-of-sample performance of the model. In this paper, we develop an extensive cross-validation scheme, inspired by $K$-fold cross-validation \citep{FriedmanESL}, that serves two purposes. First, we tune the parameters in the algorithms under study with a 5-fold cross-validation approach. Second, we evaluate the predictive performance of the algorithms investigated on multiple data sets, instead of on a single test set. Algorithm \ref{cv_algo} outlines the basic principles of our approach and Figure~\ref{cv_graph} gives a schematic representation. The full data $\mathcal{D}$ is split in six disjoint and stratified \citep{Neyman1934} subsets $\mathcal{D}_1,\ldots,\mathcal{D}_6$ by ordering first on claim frequency, then on severity. The ordered observations are assigned to each of the subsets in turn. Stratification ensures that the distribution of response variables is similar in the six subsets, as we illustrate in Table~\ref{resp_strat} for the data introduced in Section~\ref{data}. The \texttt{foreach} and inner \texttt{for} loop in Algorithm \ref{cv_algo} represent the typical approach to perform \num{5}-fold cross-validation on data from which we already separated a hold-out test set $\mathcal{D}_k$. The \texttt{foreach} loop iterates over the tuning grid and the \texttt{for} loop allows the validation set $\mathcal{D}_\ell$ to vary. The optimal tuning parameters are those that minimize the cross-validation error, which is obtained by averaging the error on the validation sets. The outer \texttt{for} loop in Algorithm \ref{cv_algo} allows the hold-out test set to vary and model performance is evaluated on this test set $\mathcal{D}_k$. Advantages of evaluating a trained model on multiple test sets are threefold. First, we obtain multiple performance measures per model class which results in a more accurate performance assessment. Second, it allows to perform sensitivity checks to assess the stability of different algorithms. Third, it exempts us from the choice of a specific test set which could bias results.

\begin{algorithm}[h]
	\SetKwInput{Input}{Input}
	\SetKwInput{Output}{Output}
	\vspace{-0mm} \hrulefill \\
	\Input{model class (\texttt{mclass}) and corresponding tuning grid (\texttt{tgrid})}
	split data $\mathcal{D}$ into $6$ disjoint stratified subsets $\mathcal{D}_1,\ldots,\mathcal{D}_6$\;
	\For{$k=1,\ldots, 6$}{
		leave out $\mathcal{D}_k$ as test set\;
		\ForEach{parameter combination in \texttt{tgrid}}{
			\For{$\ell \in \left\lbrace 1,\ldots,6\right\rbrace  \setminus k$}{
				train a model $f_{k\ell}$ of \texttt{mclass} on $\mathcal{D} \setminus \left\lbrace \mathcal{D}_k,\mathcal{D}_\ell \right\rbrace $\;
				evaluate the model performance on $\mathcal{D}_\ell$ using loss function $\mathscr{L}(\cdot,\cdot)$\;
				$\text{valid\_error}_{k\ell} \leftarrow \frac{1}{|\mathcal{D}_\ell|}\underset{i \in \mathcal{D}_\ell}{\sum} \mathscr{L}\{y_i,f_{k\ell}(\boldsymbol{x}_i)\}$\;
			}
			$\text{valid\_error}_k \leftarrow \frac{1}{5}\sum_{\ell \in \left\lbrace 1,\ldots,6\right\rbrace  \setminus k}\text{valid\_error}_{k\ell}$\;
		}
		optimal parameters from \texttt{tgrid} are those that minimize $\text{valid\_error}_k$\;
		train a model $f_k$ of \texttt{mclass} on $\mathcal{D} \setminus \mathcal{D}_k$ using the optimal parameters\;
		evaluate the model performance on $\mathcal{D}_k$ using loss function $\mathscr{L}(\cdot,\cdot)$\;
		$\text{test\_error}_{k} \leftarrow \frac{1}{|\mathcal{D}_k|}\underset{i \in \mathcal{D}_k}{\sum} \mathscr{L}\{y_i,f_{k}(\boldsymbol{x}_i)\}$\;
	}
	\Output{optimal tuning parameters + performance measure for each of the six folds.}
	\vspace{-3mm} \hrulefill \vspace{2mm}
	\caption{Cross-validation scheme for model tuning and performance evaluation.}
	\label{cv_algo}
\end{algorithm}

\begin{table}[h!]
	\centering
	\begin{tabular}{lcccccc}
		\toprule
		& $\mathcal{D}_1$ & $\mathcal{D}_2$ & $\mathcal{D}_3$ & $\mathcal{D}_4$ & $\mathcal{D}_5$ & $\mathcal{D}_6$ \\
		\midrule
		$\sum N_i/ \sum e_i$ & \num{0.1391687} & \num{0.1391433} & \num{0.1392443} & \num{0.1392213} & \num{0.1391517} & \num{0.1393045} \\
		$\sum L_i/ \sum N_i$ & \num{1296.165} & \num{1302.894} & \num{1324.667} & \num{1312.619} & \num{1330.884} & \num{1287.832} \\
		\bottomrule
	\end{tabular}
	\caption{Summary statistics of response variables in the different data subsets $\mathcal{D}_1$ to $\mathcal{D}_6$.}
	\label{resp_strat}
\end{table}

\subsection{Interpretability matters: opening the black box}
\label{interpret}
The GDPR's regime of ``algorithmic accountability'' and the resulting ``right to explanation'' highlight the vital importance of interpretable and transparent pricing models. However, machine learning techniques are often considered black boxes compared to statistical models such as GLMs. In a GLM, parameter estimates and their standard errors give information about the effect, uncertainty and statistical relevance of all variables. Such quick and direct interpretations are not possible with machine learning techniques, but this section introduces tools to gain insights from a model. A good source on interpretable machine learning is \citet{Molnar2019}. These tools are evaluated on the data used to train the optimal models, i.e.,~$\mathcal{D} \setminus \mathcal{D}_k$ for data fold $k$. A subset of the training data can be used to save computation time if needed.

\paragraph{Variable importance} Variable selection and model building is often a time consuming and tedious process with GLMs \citep{Henckaerts2018}. An advantage of tree-based techniques is their built-in variable selection strategy, making a priori design decisions less critical. Unraveling the variables that actually matter in the prediction is thus crucial. For $\ell \in\{1,\ldots,p\}$, \cite{Breiman1984} measure the importance of a specific feature $x_\ell$ in a decision tree $t$ by summing the improvements in the loss function over all the splits on $x_\ell$:
\begin{equation*}
\mathcal{I}_\ell(t) = \sum_{j=1}^{J-1} \, \mathbbm{1}\{v(j) = \ell\} \,\,\, (\Delta \mathscr{L})_j\,. 
\end{equation*}
The sum is taken over all $J-1$ internal nodes of the tree, but only the nodes where the splitting variable $x_v$ is $x_\ell$ contribute to this sum. These contributions $(\Delta \mathscr{L})_j$ represent the difference between the evaluated loss function before and after split $j$ in the tree. The idea is that important variables appear often and high in the decision tree such that the sum grows largest for those variables. We normalize these variable importance values such that they sum to 100\%, giving a clear idea about the relative contribution of each variable in the prediction.

We can easily generalize this approach to the ensemble techniques by averaging the importance of variable $x_\ell$ over the different trees that compose the ensemble:
\begin{equation*}
\mathcal{I}_\ell = \frac{1}{T} \sum_{t=1}^{T} \, \mathcal{I}_\ell(t)\,,
\end{equation*}
where the sum is taken over all trees in the random forest or gradient boosting machine.

\paragraph{Partial dependence plots} Besides knowing which variables are important, it is meaningful to understand their effect on the prediction target. Partial dependence plots, introduced in \citet{Friedman2001}, show the marginal effect of a variable on the predictions obtained from a model. Hereto, we evaluate the prediction function in specific values of the variable of interest $x_\ell$ for $\ell \in\{1,\ldots,p\}$, while averaging over a range of values of the other variables $\boldsymbol{x}^*$:
\begin{equation}
\bar{f}_\ell(x_\ell) = \frac{1}{n} \sum_{i=1}^{n} f_{\text{model}}(x_\ell,\boldsymbol{x}^*_i)\,.
\label{pdp_form}
\end{equation}
The vector $\boldsymbol{x}^*_i$ holds the realized values of the other variables for observation $i$ and $n$ is the number of observations in the training data. Interaction effects between $x_\ell$ and another variable in $\boldsymbol{x}^*$ can distort the effect \citep{Goldstein2015}. Suppose that half of the observations show a positive association between $x_\ell$ and the prediction outcome (higher $x_\ell$ leads to higher predictions), while the other half of the observations show a negative association between $x_\ell$ and the prediction outcome. Taking the average over all observations will cause the partial dependence plot to look like a horizontal line, wrongly indicating that $x_\ell$ has no effect on the prediction outcome. Individual conditional expectations can rectify such wrong conclusions.

\paragraph{Individual conditional expectation} Individual conditional expectations, introduced by \citet{Goldstein2015}, also show the effect of a variable on the predictions obtained from a model, but on an individual level. We evaluate the prediction function in specific values of the variable of interest $x_\ell$ for $\ell \in\{1,\ldots,p\}$, keeping the values of the other variables $\boldsymbol{x}^*$ fixed:
\begin{equation}
\tilde{f}_{\ell,i}(x_\ell) = f_{\text{model}}(x_\ell,\boldsymbol{x}^*_i)\,,
\label{ice_form}
\end{equation}
where $\boldsymbol{x}^*_i$ are the realized values of the other variables for observation $i$.  We obtain an effect for each observation $i$, allowing us to detect interaction effects when some (groups of) observations show different behavior compared to others. For example, two distinct groupings will emerge when half of the observations have a positive association and the other half a negative association between $x_\ell$ and the prediction outcome. Individual conditional expectations can also be used to investigate the uncertainty of the effect of variable $x_\ell$ on the prediction outcome. The partial dependence plot can be interpreted as the average of this collection of individual conditional expectations, i.e.,~$\bar{f}_\ell(x_\ell) = \frac{1}{n} \sum_{i=1}^{n} \tilde{f}_{\ell,i}(x_\ell) $.

\section{Case study: claim frequency and severity modeling}
\label{ResultFreqSev}
An insurer's pricing team uses proprietary data to deliver a fine-grained tariff plan for a portfolio. As a typical example of such data, we study a motor third party liability (MTPL) portfolio from a Belgian insurer in 1997. This section puts focus on the claim frequency and severity models that are developed with the different modeling techniques. We briefly introduce the data and we report the optimal tuning parameters for the frequency and severity models. Afterwards, we use the tools from Section~\ref{interpret} to gain some insights in these optimal models. We conclude this section with an out-of-sample deviance comparison to assess the statistical performance of the different modeling techniques.

\subsection{Quick scan of the MTPL data}
\label{data}
The data used here is also analyzed in \citet{Denuit2004}, \citet{Klein2014} and \citet{Henckaerts2018}. We follow the same data pre-processing steps as the aforementioned papers, e.g.,~regarding the exclusion of very large claims. Table~\ref{variables} in Appendix \ref{App_var} lists a description of the available variables. The portfolio contains \num{163212} unique policyholders, each observed during a period of exposure-to-risk expressed as the fraction of the year during which the policyholder was exposed to the risk of filing a claim. Claim information is known in the form of the number of claims filed and the total amount claimed (in euro) by a policyholder during her period of exposure. The data set lists five categorical, four continuous and two spatial risk factors, each of them informing about specific characteristics of the policy or the policyholder. A detailed discussion on the distribution of all variables is available in \citet{Henckaerts2018}. Regarding spatial information, we have access to the 4-digit postal code of the municipality of residence and the accompanying latitude/longitude coordinates of the center of this area. The GAM/GLM benchmarks employ spatial smoothing over the latitude/longitude coordinates. In line with this approach, we use the coordinates as continuous variables in the tree-based models. 

\begin{figure}[h!]
	\includegraphics[width=0.94\textwidth]{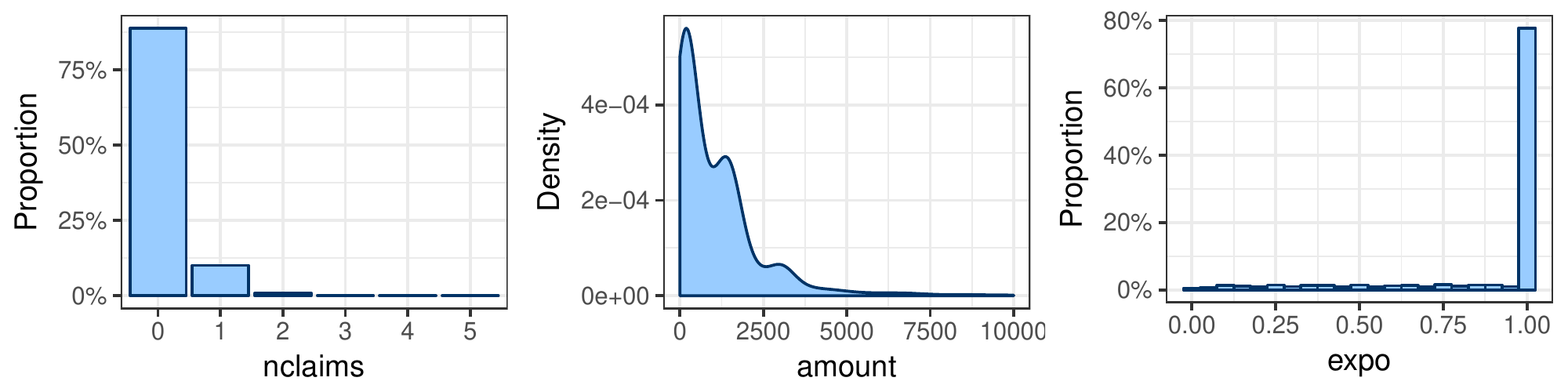}
	\centering
	\caption{Distribution of the claim counts, amounts and the exposure-to-risk measure in the \texttt{MTPL} data.}
	\label{claims_expo}
\end{figure}

Figure~\ref{claims_expo} displays the distribution of the claims information (\texttt{nclaims} and \texttt{amount}) and the exposure-to-risk measure (\texttt{expo}). Most policyholders in the portfolio are claim-free during their insured period, some file one claim and few policyholders file two, three, four or five claims. The majority of all these claims involve small amounts, but very large claims occur as well. Most policyholders are exposed to the risk during the full year, but there are policyholders who started the policy in the course of the year or surrendered the policy before the end of the year. Figure~\ref{claims_expo} motivates the use of loss functions which are not based on the squared error loss. We work with the Poisson and gamma distribution/deviance for frequency and severity respectively as our preferred distributional assumption (for GAM/GLM) and loss function choice (for tree-based techniques). Note that earlier work on this data, such as \citet{Denuit2007} and \citet{Henckaerts2018}, assumed a log-normal distribution for severity. We illustrate the difference and motivate our choice in the supplementary material.

\subsection{Optimal tuning parameters}
Table~\ref{opt_params} lists the optimal tuning parameters for the different machine learning techniques in each of the six data folds. Comparing the number of splits in the trees and the number of trees in the ensembles, we conclude that the frequency models are more extensive compared to the severity variants. This is driven by the lower sample size for severity modeling and the fact that the severity of a claim is typically much harder to predict than the frequency \citep{Charpentier2014}.

The complexity parameter $cp$ does not give much information about the size of a regression tree and therefore Table~\ref{opt_params} also lists the number of splits in the tree. All frequency trees contain between \num{20} and \num{38} splits while the severity trees comprise of only one or two splits. The coefficient of variation for the gamma prior $\gamma$ remains stable over the different data folds.

The number of trees $T$ in the random forest is very unstable over the different folds for both frequency and severity. This shows that the size of the eventual model highly depends on the training data when simply averaging independently grown trees. In four out of six cases, the number of split candidates $m$ is equal to 5 for frequency models and 2 for the severity models. The low value of $m$ for the severity random forests indicates that the variance reduction is the main driver for reducing the loss, as opposed to finding the best split out of multiple candidates.

The number of trees $T$ in the gradient boosting machine is more stable over the folds compared to the random forest. This shows that the sequential approach of growing a boosted model is less affected by the specific data fold. The tree depth $d$ ranges from 3 to 5 in the frequency models. Table~\ref{opt_params} reveals that the largest values of $T$ correspond to the smallest values of $d$ and vice versa, highlighting the interplay between these tuning parameters. In five out of six cases, the severity models use stumps (i.e.,~trees with only one split) as weak learners. A tree depth of $d=1$ makes these models completely additive, without interaction by construct.

\begin{table}[h!]
\centering
\footnotesize
\begin{tabular}{@{\extracolsep{10pt}}lcccccccc@{}}
	\toprule
	&& \multicolumn{3}{c}{Regression tree} & \multicolumn{2}{c}{Random forest}  & \multicolumn{2}{c}{Boosting machine} \\
	\cmidrule(lr){3-5} 	\cmidrule(lr){6-7} 	\cmidrule(lr){8-9}
	& data fold & $cp$ & $\gamma$ & splits & $T$ & $m$ & $T$ & $d$ \\
	\midrule
	\multirow{6}{*}{Frequency} & 1 & $7.3 \times 10^{-5}$ & 0.125 & 38 & \num{4900} & 5 & \num{2600} & 3\\
	& 2 & $1.4 \times 10^{-4}$ & 0.125 & 24 & \num{900} & 5 & \num{2000} & 4\\
	& 3 & $1.1 \times 10^{-4}$ & 0.125 & 31 & \num{400} & 8 & \num{1400} & 5\\
	& 4 & $1.2 \times 10^{-4}$ & 0.250 & 27 & \num{5000} & 5 & \num{1500} & 5\\
	& 5 & $1.8 \times 10^{-4}$ & 0.250 & 20 & \num{600} & 10 & \num{1900} & 4\\
	& 6 & $1.7 \times 10^{-4}$ & 0.250 & 23 & \num{100} & 5 & \num{2700} & 3\\
	\midrule
	\multirow{6}{*}{Severity} & 1 & $3.3 \times 10^{-3}$ & - & 2 & \num{4300} & 2 & \num{600} & 1\\
	& 2 & $5.8 \times 10^{-3}$ & - & 1 & \num{200} & 2 & \num{300} & 1\\
	& 3 & $3.7 \times 10^{-3}$ & - & 2 & \num{600} & 1 & \num{500} & 1\\
	& 4 & $7.3 \times 10^{-3}$ & - & 1 & \num{100} & 2 & \num{400} & 2\\
	& 5 & $5.4 \times 10^{-3}$ & - & 1 & \num{3600} & 2 & \num{600} & 1\\
	& 6 & $5.4 \times 10^{-3}$ & - & 1 & \num{100} & 1 & \num{600} & 1\\
	\bottomrule
\end{tabular}
	\caption{Overview of the optimal tuning parameters for the tree-based machine learning techniques.}
	\label{opt_params}
\end{table}

We also tune the benchmark GLMs for each of the six data folds separately, i.e.,~we perform the binning strategy from \citet{Henckaerts2018} in each fold $k$ such that the optimal bins are chosen for the training data at hand $\mathcal{D} \setminus \mathcal{D}_k$. A grid is used for the two tuning parameters involved, one for the continuous variables and one for the spatial effect, thereby avoiding the two-step binning procedure initially proposed in \citet{Henckaerts2018}. Examples of the resulting benchmark GLMs for frequency and severity are presented in Appendix \ref{App_glm}.

%

\subsection{Model interpretation}
\label{modinterpret}

We will use the variable importance measure to find the most relevant variables in the frequency and severity models. Afterwards, we will make use of partial dependence plots and individual conditional expectations to gain understanding on a selection of interesting effects for the claim frequency. Similar results on claim severity can be found in the supplementary material.

\begin{figure}[h!]
	\centering
	\includegraphics[width=0.94\textwidth]{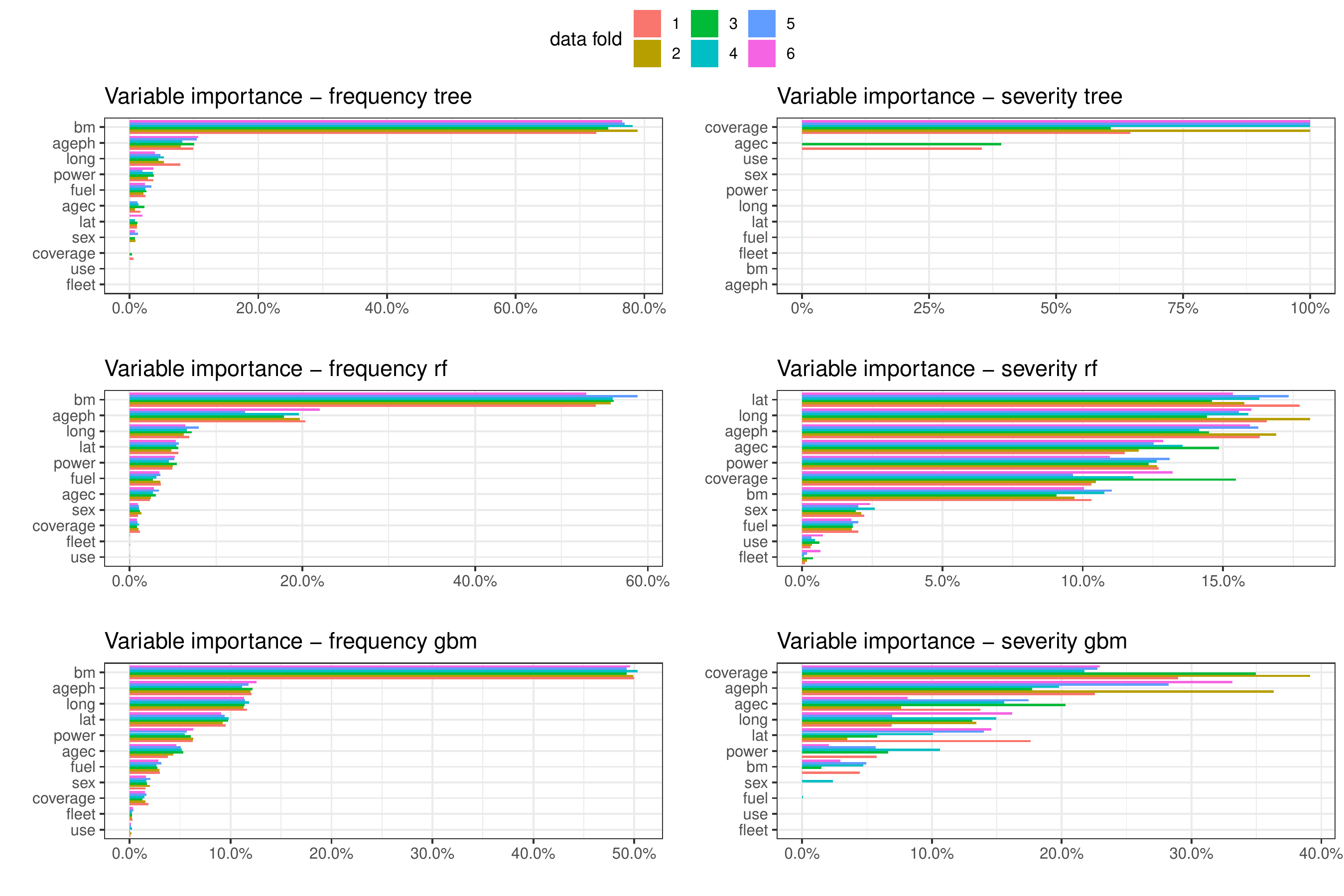}
	\caption{Variable importance in the optimal regression tree (top), random forest (middle) and gradient boosting machine (bottom) per data fold (color) for frequency (left) and severity (right).}
	\label{var_imp}
\end{figure}

\paragraph{Variable importance} To learn which variables matter for predicting claim frequency and severity, we compare in Figure~\ref{var_imp} the variable importance plots for the different machine learning techniques over the six data folds. The variables are ranked from top to bottom, starting with the most important one as measured by the average variable importance over the folds (multiple variables with zero importance are ordered alphabetically from Z to A). By contrasting the graphs in the left column of Figure~\ref{var_imp}, we see that the important variables (mostly bonus-malus scale and age) are similar across all methods for the frequency model. Other relevant risk factors are the power of the vehicle and the spatial risk factor (combining the longitude and latitude information). The frequency GLM, presented in Table~\ref{glm_freq_data3} in Appendix~\ref{App_glm}, contains the top seven variables together with \texttt{coverage}, which is ranked at the ninth place for all methods.

The right column of Figure~\ref{var_imp} shows the variable importance for severity models. The ranking is very dissimilar across the different modeling techniques. The regression tree models for severity contain only one split using the type of coverage in four out of the six folds, while the other two trees have an additional split on the age of the car. The random forest and gradient boosting machine include more variables, but they both lead to rather different rankings of the importance of those variables. The severity GLM, presented in Table~\ref{glm_sev_data3} in Appendix~\ref{App_glm}, contains three variables: \texttt{coverage}, \texttt{ageph} and \texttt{agec}. An interesting observation is that the most important risk factors in the gradient boosting machines are those selected in the GLMs.

\begin{figure}[h!]
	\centering
	\includegraphics[width=0.94\textwidth]{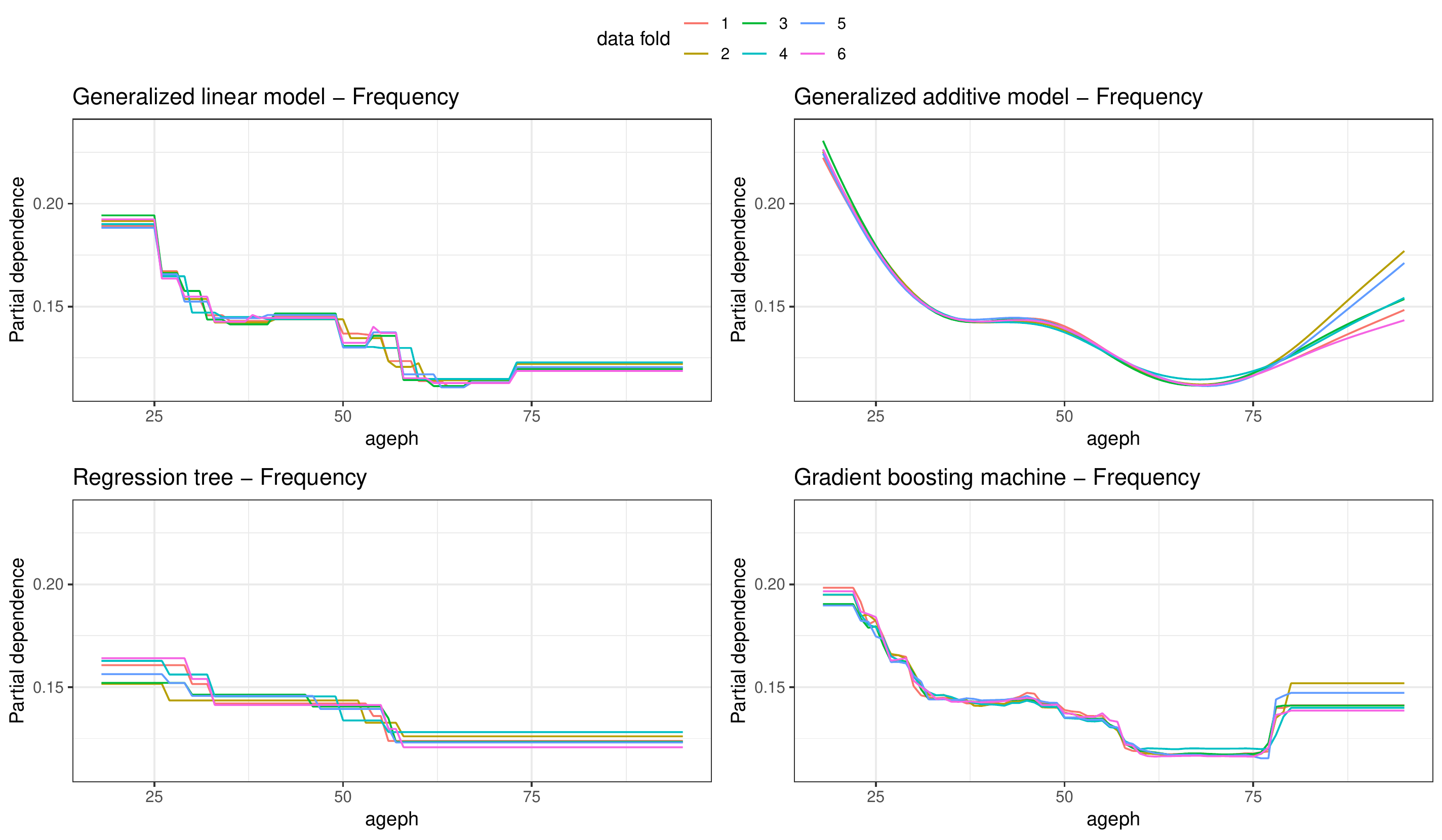}
	\caption{Partial dependence plot to visualize the effect of the age of the policyholder on frequency for the optimal model obtained per data fold (color) in a GLM (top left), GAM (top right), regression tree (bottom left) and gradient boosting machine (bottom right).}
	\label{pdp_freq_ageph}
\end{figure}

\paragraph{Partial dependence plot of the age effect} Figure~\ref{pdp_freq_ageph} compares the partial dependence effect of the age of the policyholder in frequency models. The two top panels of Figure~\ref{pdp_freq_ageph} show the GLM and GAM effects on the left and right respectively. As explained in Section~\ref{classicpriceframe}, due to our proposed data-driven approach, the GLM effects are a step-wise approximation of the GAM effects. The risk of filing a claim is high for young policyholders and gradually decreases with increasing ages to stabilize around the age of 35. The risk starts decreasing again around the age of 50 and increases for senior policyholders around the age of 70. The bottom left panel of Figure~\ref{pdp_freq_ageph} shows the age effect captured by the regression trees. The effect is less stable across the folds compared to the other methods, this is a confirmation and illustration of the variability of a single regression tree. There is also no increase in risk for senior policyholders in the regression trees. The bottom right panel of Figure~\ref{pdp_freq_ageph} shows the age effect according to the gradient boosting machines. This looks very similar to the smooth GAM effect with one important distinction, namely the flat regions at the boundaries. This makes the tree-based techniques more robust with respect to extrapolation and results in less danger of creating very high premiums for risk profiles at edges. Note that the gradient boosting machine predicts a wider range of frequencies than the regression tree, namely $\num{0.12}$ to $\num{0.20}$ versus $\num{0.12}$ to $\num{0.165}$ respectively. The shape of the age effect in the random forest, available in Appendix~\ref{rf_extra}, is rather similar to the gradient boosting machine effect but on a slightly more compact range. 

\begin{figure}[h!]
	\centering
	\includegraphics[width=0.94\textwidth]{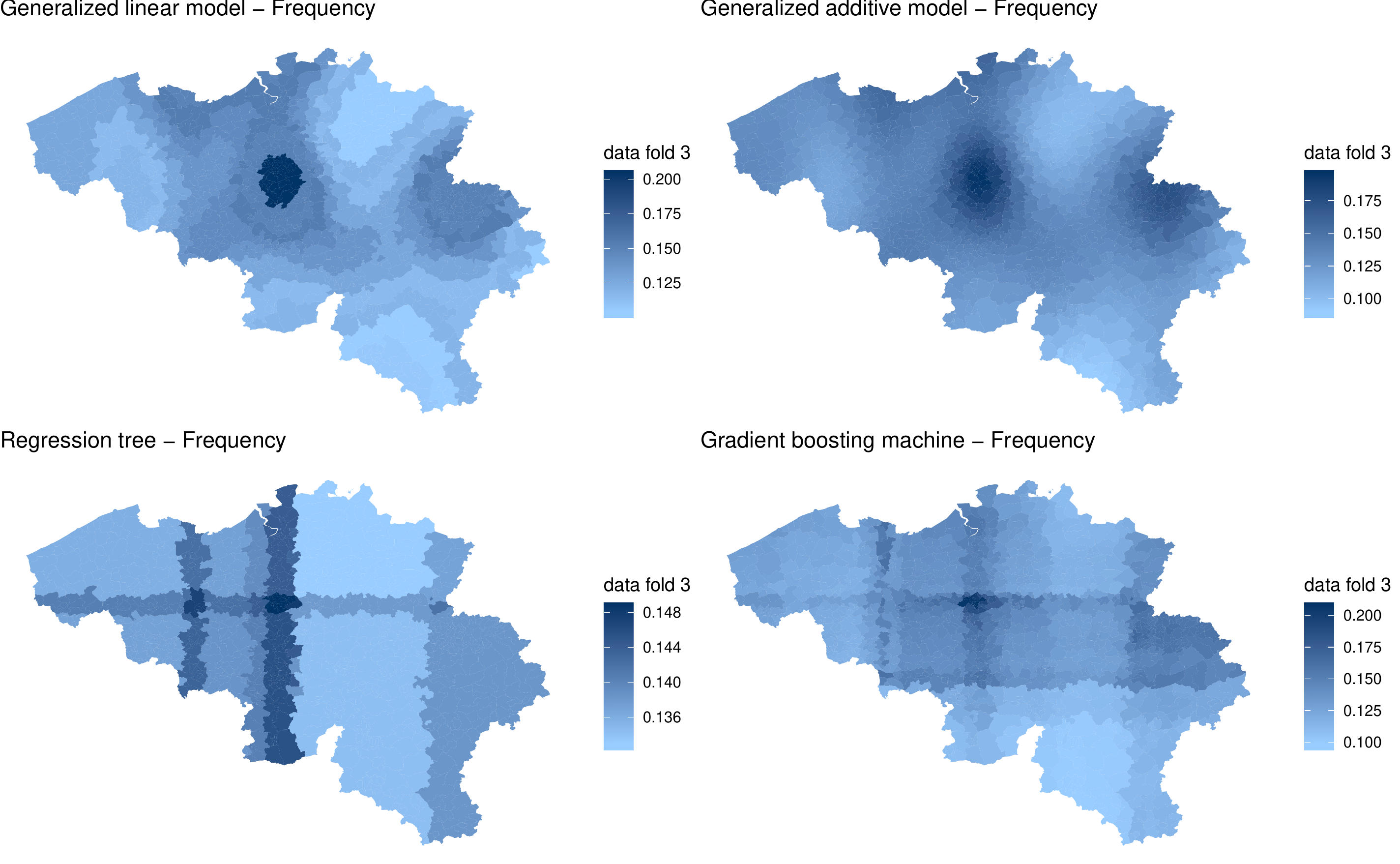}
	\caption{Partial dependence plot to visualize the effect of the municipality of residence on frequency in a GLM (top left), GAM (top right), regression tree (bottom left) and gradient boosting machine (bottom right).}
	\label{pdp_freq_spatial}
\end{figure}

\paragraph{Partial dependence plot of the spatial effect} Figure~\ref{pdp_freq_spatial} compares the spatial effect in frequency models, more specifically the models trained on the data where fold $\mathcal{D}_3$ was kept as the hold-out test set. We choose a specific data fold because we otherwise need to show six maps of Belgium per method as opposed to overlaying six effects as in Figure~\ref{pdp_freq_ageph}. The two top panels show the GLM and GAM effects on the left and right respectively. Brussels, the capital of Belgium located in the center of the country, is clearly the most accident-prone area to live and drive a car because of heavy traffic. The southern and northeastern parts of Belgium are less risky because of sparser population and more rural landscapes. The bottom left panel of Figure~\ref{pdp_freq_spatial} shows the spatial effect as it is captured with a regression tree. Splitting on longitude and latitude coordinates results in a rectangular split pattern on the map of Belgium. The bottom right panel of Figure~\ref{pdp_freq_spatial} shows the spatial effect for the gradient boosting machine. The underlying rectangular splits are still visible but in a smoother way compared to the regression tree. Brussels still pops out as the most risky area and the pattern looks similar to the GLM and GAM effects. The shape of the spatial effect in the random forest, available in Appendix~\ref{rf_extra}, is again similar to that of the gradient boosting machine on a more compact range.

Figures~\ref{pdp_freq_ageph} and \ref{pdp_freq_spatial} teach us that a single tree is not able to capture certain aspects in the data, resulting in a coarse approximation of the underlying risk effect. The ensemble techniques are able to capture refined risk differences in a much smoother way. The differences in the range of the predicted effects imply that the gradient boosting machine performs more segmentation while the random forest puts more focus on risk pooling.

\begin{figure}[h!]
	\centering
	\includegraphics[width=0.94\textwidth]{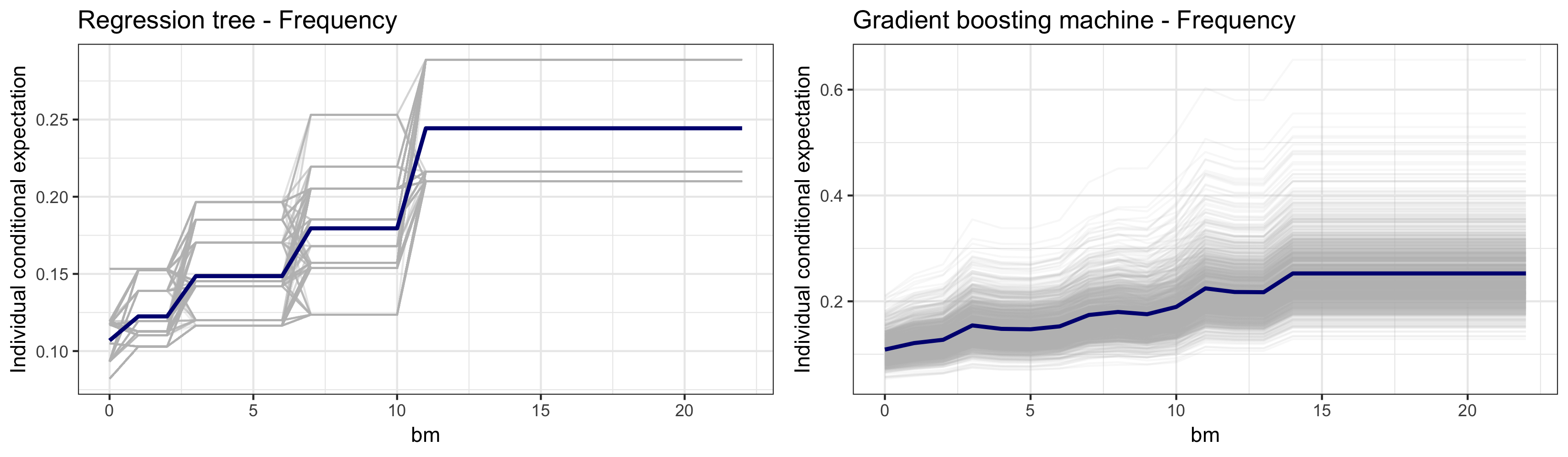}
	\caption{Effect of the bonus-malus scale on frequency in a regression tree (left) and gradient boosting machine (right) as partial dependence (blue) and individual conditional expectations (gray).}
	\label{ice_freq_bm}
\end{figure}

\paragraph{Individual conditional expectation of the bonus-malus effect} Figure~\ref{ice_freq_bm} compares the bonus-malus effect captured with a regression tree (left) and a gradient boosting machine (right) for frequency data. As in Figure~\ref{pdp_freq_spatial}, we show the effect for the models trained on the data where fold $\mathcal{D}_3$ was kept as the hold-out test set. The gray lines are individual conditional expectations for \num{1000} random policyholders and the blue line shows the partial dependence curve. The values for $\boldsymbol{x}^*$ in Eq.~\eqref{ice_form} are those registered for the selected policyholders. On average, we observe an increase in frequency risk as the blue line surges over the bonus-malus levels, which is to be expected because higher bonus-malus levels indicate a worse claim history. We can get an idea about the sensitivity of the bonus-malus effect across the different policyholders in the portfolio by comparing the steepness of the gray lines. Keeping all other risk factors fixed, a steeper effect indicates that a policyholder's risk is more sensitive to changes in the bonus-malus scale. This effect is driven by the combination of all risk factors registered for this policyholder. 

The previous figures show some counterintuitive results regarding the monotonicity of a fitted effect. For example, the bonus-malus individual conditional expectations for the regression tree in Figure~\ref{ice_freq_bm} reveal decreases in risk over increasing bonus-malus levels for certain policyholders. This poses problems for the practical implementation of such a tariff because it assigns a lower premium to policyholders with a worse claim history. In practice, an actuary would specify monotonicity constraints on such a risk factor, either by an a posteriori smoothing of the resulting effect or by using an implementation that allows to specify such constraints a priori, e.g.,~the \texttt{gbm} package has this functionality. Our analysis does not enforce such constraints.

\subsection{Hunting for interaction effects}
Tree-based models are often praised for their ability to model interaction effects between variables \citep{Buchner2017,Schiltz2018}. The predictions of a model can not be expressed as the sum of the main effects when interactions are present, because the effect of one variable depends on the value of another variable. Friedman's $H$-statistic, introduced by \citet{Friedman2008}, estimates the interaction strength by measuring how much of the prediction variance originates from the interaction effect. We will put focus on two-way interactions between variables $x_k$ and $x_\ell$, but in theory this measure can be applied to arbitrary interactions between any number of variables. Let $\bar{f}_k(x_k)$ and $\bar{f}_l(x_\ell)$ represent the one-dimensional partial dependence of the variables as defined in Section~\ref{interpret} and $\bar{f}_{kl}(x_k,x_\ell)$ the two-way partial dependence, defined analogously to Eq.~\eqref{pdp_form}. The $H$-statistic is expressed as:
\begin{equation*}
H^2_{k\ell} = \frac{\sum_{i=1}^{n}\{ \bar{f}_{kl}(x^{(i)}_k,x^{(i)}_\ell) - \bar{f}_k(x^{(i)}_k) - \bar{f}_l(x^{(i)}_\ell)  \}^2 }{\sum_{i=1}^{n} \bar{f}_{kl}^2(x^{(i)}_k,x^{(i)}_\ell)} \, ,
\end{equation*}

where $x^{(i)}_k$ indicates that the partial dependence function is evaluated at the observed value of $x_k$ for policyholder $i$. Assuming the partial dependence is centered at zero, the numerator measures the variance of the interaction while the denominator measures the total variance. The ratio of both therefore measures the interaction strength as the amount of variance explained by the interaction. The $H$-statistic ranges from zero to one, where zero indicates no interaction and one implies that the effect of $x_k$ and $x_\ell$ on the prediction is purely driven by the interaction.

Table~\ref{hstat} shows the fifteen highest two-way $H$-statistics among the variables available in the data set (as listed in Table~\ref{variables} in Appendix~\ref{App_var}) for the frequency gradient boosting machine trained on the data where fold $\mathcal{D}_3$ was kept as the hold-out test set. The strongest interaction is found between the longitude and latitude coordinates, which is not a surprise seeing how these two variables together encode the region where the policyholder resides. 

\begin{table}[h!]
	\centering
	\setlength\tabcolsep{5pt}
	\begin{tabular}{lc|lc|lc}
		\toprule
		Variables & $H$-statistic & Variables & $H$-statistic & Variables & $H$-statistic \\
		\midrule
		(\texttt{lat}, \texttt{long})& \num{0.2687} & \cellcolor{gray!25}(\texttt{agec}, \texttt{coverage}) & \cellcolor{gray!25}\num{0.1185} & \cellcolor{gray!25}(\texttt{bm}, \texttt{power}) & \cellcolor{gray!25}\num{0.0800} \\
		\cellcolor{gray!25}(\texttt{fuel}, \texttt{power})& \cellcolor{gray!25}\num{0.1666} & \cellcolor{gray!25}(\texttt{ageph}, \texttt{power}) & \cellcolor{gray!25}\num{0.1062} & (\texttt{ageph}, \texttt{lat}) & \num{0.0799} \\
		(\texttt{agec}, \texttt{power})& \num{0.1319} & (\texttt{ageph}, \texttt{bm}) & \num{0.0961} & \cellcolor{gray!25}(\texttt{agec}, \texttt{ageph}) & \cellcolor{gray!25}\num{0.0785}  \\
		\cellcolor{gray!25}(\texttt{ageph}, \texttt{sex})& \cellcolor{gray!25}\num{0.1293}  & (\texttt{power}, \texttt{sex}) & \num{0.0829} &  (\texttt{long}, \texttt{sex})& \num{0.0732} \\
		(\texttt{coverage}, \texttt{long})& \num{0.1203} & (\texttt{fuel}, \texttt{long}) & \num{0.0828} & (\texttt{agec}, \texttt{bm}) & \num{0.0678} \\
		\bottomrule
	\end{tabular}
	\caption{$H$-statistic of the 15 strongest two-way interactions between all the variables in the gradient boosting machine for frequencies, trained on data with $\mathcal{D}_3$ as hold-out test set.}
	\label{hstat}
\end{table}

\begin{figure}[h!]
	\centering
	\includegraphics[width=0.94\textwidth]{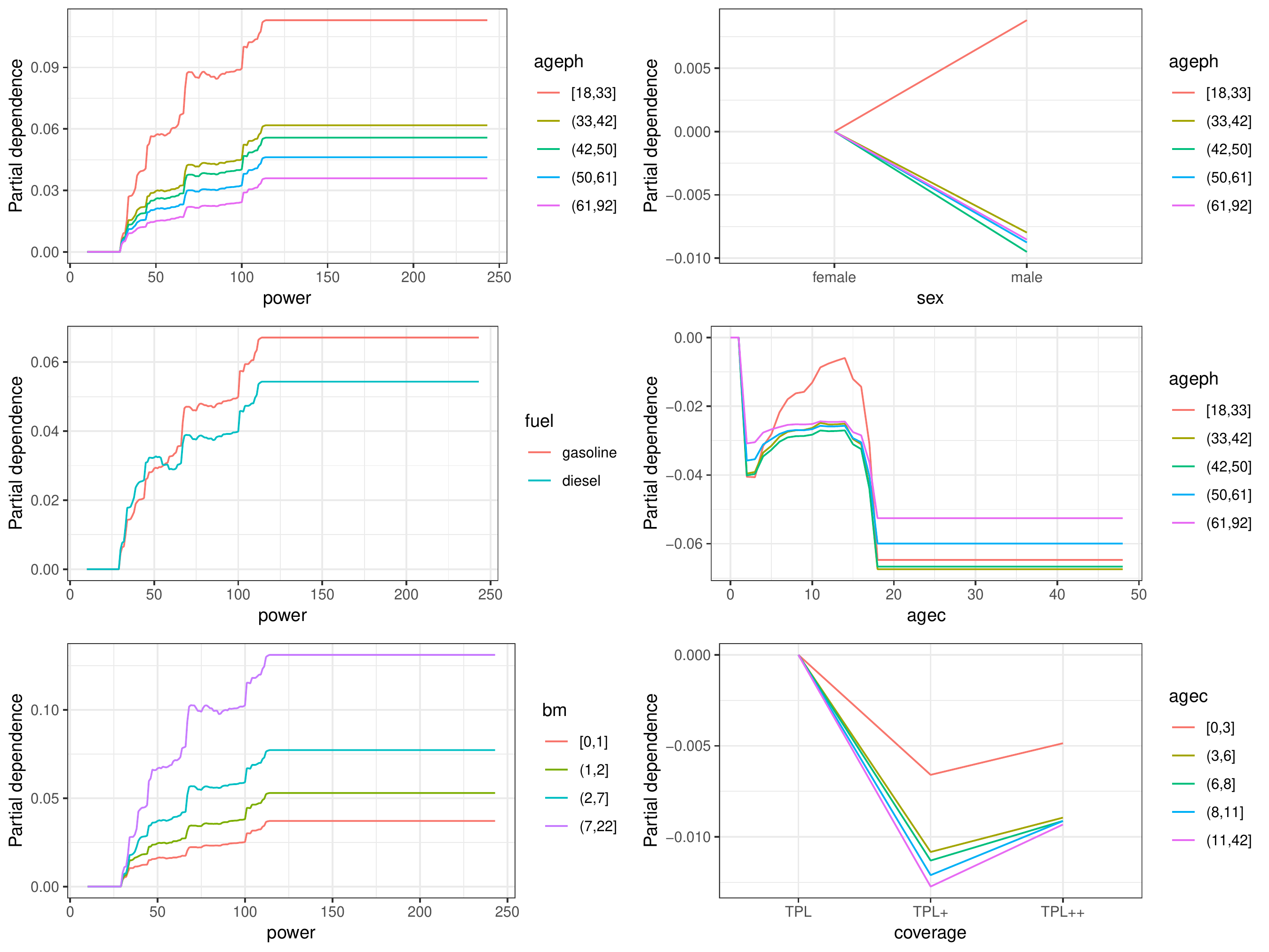}
	\caption{Grouped partial dependence plots for the gradient boosting machine on frequency, trained on data with $\mathcal{D}_3$ as hold-out test set. The effect is binned in five equally sized groups. The left column shows the effects for the power of the car grouped by the age of the policyholder (top), the type of fuel (middle) and the bonus-malus scale (bottom). The right column shows the effects for the sex of the policyholder (top), age of the car (middle) and type of coverage (bottom), grouped by the age of the policyholder or car.}
	\label{pdp_interactions}
\end{figure}

The $H$-statistic informs us on the strength of the interaction between two variables, but gives us no idea on how the effect behaves. Figure~\ref{pdp_interactions} shows grouped partial dependence plots to investigate the interactions highlighted in gray in Table~\ref{hstat}. The partial dependence plots of a specific variable are grouped into five equally sized groups based on the value of another variable. Interaction effects between both variables can be discovered by comparing the evolution of the curves over the five different groups. An interaction is at play when this evolution is different for policyholders in different groups. In order to focus purely on the evolution of the effect, we let all the curves in Figure~\ref{pdp_interactions} start at zero by applying a vertical shift.

An important and well-known effect in insurance pricing is the interaction between the age of the policyholder and the power of the vehicle. Our benchmark GLM and GAM use this interaction effect in the predictor and Figure 11 in \cite{Henckaerts2018} shows that young policyholders with high power vehicles form an increased risk for the insurer regarding claim frequency. The top left panel of Figure~\ref{pdp_interactions} shows the partial dependence of the power of the vehicle, grouped by the age of the policyholder. The power effect is steepest for young policyholders, indicated by the red line. The steepness of the effect decreases for increasing ages. The difference in the steepness of the effect between young and old policyholders is a visual confirmation of the interaction at play between the variables \texttt{ageph} and \texttt{power}. The top right panel of Figure~\ref{pdp_interactions} shows the partial dependence for the sex of the policyholders, grouped by their age. For young policyholders, aged 18 to 33, we observe that males are on average more risky drivers compared to females, while for the other age groups the female drivers are perceived more risky than males. European insurers are not allowed to use gender in their tariff structure nowadays, implying that young female drivers might be partly subsidizing their male peers.

The middle left panel of Figure~\ref{pdp_interactions} shows the partial dependence of the power of the vehicle, grouped by the type of fuel. We observe that the steepness of the power effect is slightly higher for gasoline cars. Drivers typically choose a diesel car when their annual mileage is above average, which would justify their choice of buying a bigger and more comfortable car with higher horsepower. However, drivers who own a high powered gasoline car might choose such a car to accommodate for a more sportive driving style, making them more prone to the risk of a claim. The middle right panel of Figure~\ref{pdp_interactions} shows the partial dependence of the age of the vehicle, grouped by the policyholder's age. We observe a big bump for young policyholders in the vehicle age range from 5 to 15. This could indicate an increased claim frequency risk for starting drivers who buy their first car on the second-hand market. The sharp drop around 19 could relate to vintage cars that are used less often and are thus less exposed to the claim risk.

The bottom left panel of Figure~\ref{pdp_interactions} shows the partial dependence of the power of the vehicle, grouped by the bonus-malus scale. We observe that the power effect grows steeper for increasing levels occupied in the bonus-malus scale. This indicates that driving a high powered car becomes more risky for policyholders in higher bonus-malus scales. The bottom right panel of Figure~\ref{pdp_interactions} shows the partial dependence of the type of coverage, grouped by the age of the vehicle. For vehicles in the age range zero to three, we observe that adding material damage covers decreases the claim frequency risk less compared to other age ranges. This might indicate that policyholders who buy a new car add a material damage cover because they worry about damaging their newly purchased vehicle, while policyholders with older cars who still add damage covers are more risk-averse and also less risky drivers.

\subsection{Statistical out-of-sample performance}
\label{oos_comp}
\begin{figure}[h!]
	\centering
	\includegraphics[width=0.94\textwidth]{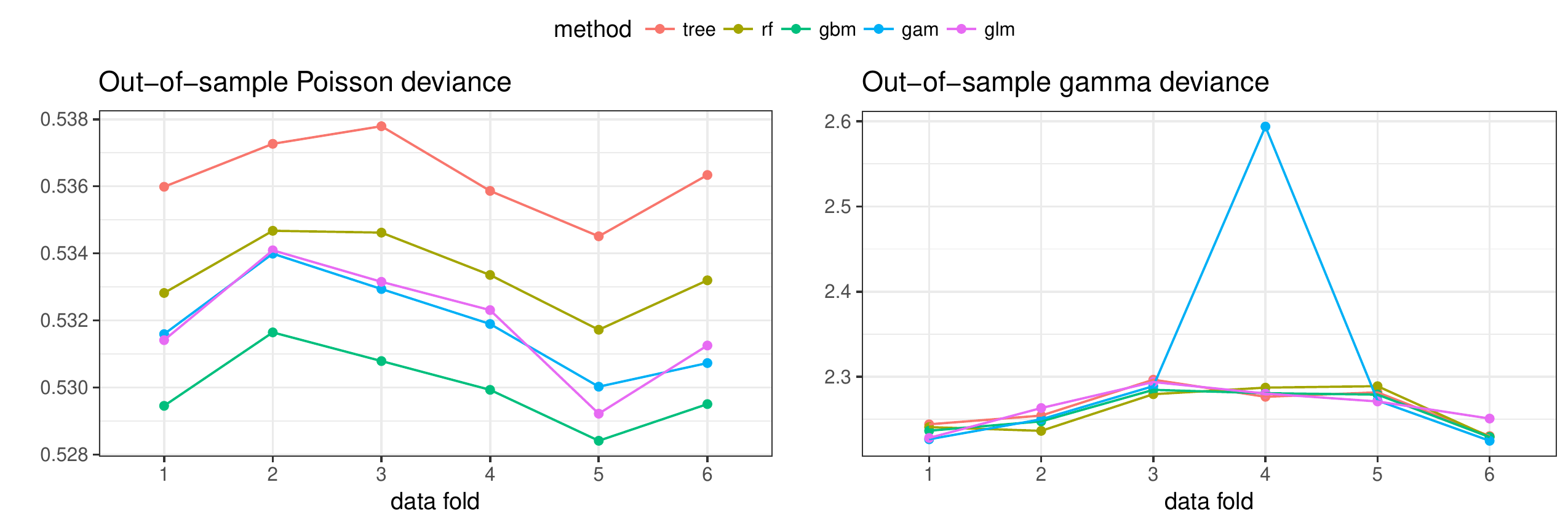}
	\caption{Out-of-sample Poisson deviance for frequency (left) and gamma deviance for severity (right), each color representing a different modeling technique.}
	\label{oos_dev}
\end{figure}

Figure~\ref{oos_dev} compares the out-of-sample performance of the different models investigated over the six data folds. We evaluate the Poisson deviance for frequency models and the gamma deviance for severity models, see Eq.~\eqref{poiss_dev} and \eqref{gamma_dev} respectively, on the holdout test data. In the left panel of Figure~\ref{oos_dev}, we observe a clear ranking of the out-of-sample Poisson deviance among the different methods. 
The gradient boosting machine is most predictive, consistently leading to the lowest deviance values. The performance of GLMs and GAMs is very similar, which is expected because the GLM is a data driven approximation of the GAM, as explained in Section~\ref{classicpriceframe}. Next in line is the random forest, and the regression tree is the least predictive for frequency. These results are very stable over the six data folds. The right panel of Figure~\ref{oos_dev} shows the out-of-sample gamma deviance for severity. The methods perform rather similarly and there is no clear winner or loser over the different folds. The peak at the fourth fold reveals a weakness of the GAM: the extrapolation of smooth effects, see Figure~\ref{pdp_freq_ageph}, combined with out-of-sample testing can lead to huge deviance values. In the severity GAM trained on $\mathcal{D} \setminus \mathcal{D}_4$ and evaluated on $\mathcal{D}_4$, the problem occurred with the age of the car. Specifically, the maximal value for \texttt{agec} in $\mathcal{D} \setminus \mathcal{D}_4$ for severity is 32 while the maximal value in $\mathcal{D}_4$ is 37, thus requiring an extrapolation of the calibrated smooth effect. This motivates to cap continuous variables at a certain cut-off in a pre-processing stage for a GAM. A tree-based method automatically deals with this problem thanks to the flat region at the outer ends of the effect, see Figure~\ref{pdp_freq_ageph}.

This comparison only puts focus on the statistical performance of the frequency and severity models. In the next section, we combine both in a pricing model and compare the different tariff structures using practical business metrics relevant for an insurance company.
\section{Model lift: from analytic to managerial insights}
\label{business}
Choosing between two tariff structures is an important business decision. This creates the need to translate our model findings to evaluation criteria that capture a manager's interest. We evaluate the economic value of a tariff using tools proposed in the literature to measure model lift \citep[Section 7.2]{Frees2013,Goldburd2016}. In this context, model lift refers to the ability of a model to prevent adverse selection. An insurer might become victim hereof when a competitor refines its tariff structure via innovation such that good risks switch to the competitor and the insurer is left with the bad risks which are more prone to high losses.

We combine the claim frequency and severity models from Section~\ref{ResultFreqSev} to obtain the technical premium for each policy under consideration, allowing us to compare different tariff structures. As Figure~\ref{cv_graph} illustrates, each observation is out-of-sample in exactly one of the data folds, more specifically the observations in $\mathcal{D}_k$ are out-of-sample for data fold $k$. We then use the optimal model trained on $\mathcal{D} \setminus \mathcal{D}_k$ to predict the policyholders in holdout test set $\mathcal{D}_k$. Following this strategy, we obtain one premium per modeling technique for each policyholder in the full data~$\mathcal{D}$. Table~\ref{prem_vs_loss} shows a comparison between the predicted premium totals and the observed losses, both on the portfolio level and split by data fold. On average, every method is perfectly capable of replicating the total losses. Section~\ref{adv_sel} compares the model lift measures from Section~\ref{modlift_tools} on the portfolio level. We also analyzed each of the data folds separately (not shown), leading to the same ranking of models as in Section~\ref{adv_sel}, thereby validating the consistency of our results.

\begin{table}[h!]  \small
	\begin{subtable}{0.72\textwidth}
		\setlength\tabcolsep{4pt}
		\begin{tabular}{cSSSS|S} 
			\toprule
			Data fold& \multicolumn{1}{c}{GLM} & \multicolumn{1}{c}{CART} & \multicolumn{1}{c}{RF} & \multicolumn{1}{c}{GBM} & \multicolumn{1}{c}{Losses}\\ 
			\midrule
			1 & \num{4396698} & \num{4341397} & \num{4407389} & \num{4376619} & \num{4365483} \\ 
			2 & \num{4420933} & \num{4339615} & \num{4419903} & \num{4384513} & \num{4388147}\\ 
			3 & \num{4369876} & \num{4313768} & \num{4380972} & \num{4337848} & \num{4461478}\\ 
			4 & \num{4370502} & \num{4374666} & \num{4383748} & \num{4324014} & \num{4422213}\\ 
			5 & \num{4405369} & \num{4374368} & \num{4399226} & \num{4357937} & \num{4485079}\\ 
			6 & \num{4397372} & \num{4375151} & \num{4412852} & \num{4363588} & \num{4342569}\\
			\midrule 
			portfolio & \num{26360750} & \num{26118966} & \num{26404091} & \num{26144519} & \num{26464970} \\ 
			\bottomrule
		\end{tabular}
	\end{subtable}%
	\begin{subtable}{0.3\textwidth}
		\setlength\tabcolsep{4pt}
		\begin{tabular}{SSSS} 
			\toprule
			\multicolumn{1}{c}{GLM} & \multicolumn{1}{c}{CART} & \multicolumn{1}{c}{RF} & \multicolumn{1}{c}{GBM} \\ 
			\midrule
			\num{1.01} & \num{0.99} & \num{1.01} & \num{1.00}  \\ 
			\num{1.01} & \num{0.99} & \num{1.01} & \num{1.00} \\ 
			\num{0.98} & \num{0.97} & \num{0.98} & \num{0.97} \\ 
			\num{0.99} & \num{0.99} & \num{0.99} & \num{0.98} \\ 
			\num{0.98} & \num{0.98} & \num{0.98} & \num{0.97}\\ 
			\num{1.01} & \num{1.01} & \num{1.02} & \num{1.01}\\
			\midrule 
			\num{1.00} & \num{0.99} & \num{1.00} & \num{0.99} \\ 
			\bottomrule
		\end{tabular}
	\end{subtable}
	\caption{Comparison of the predicted premiums and observed losses on portfolio level and by data fold. We show the premium and loss totals (left) and the ratio of premiums to losses (right).} 
	\label{prem_vs_loss} 
\end{table}

\subsection{Tools to measure model lift}
\label{modlift_tools}
Suppose that an insurance company has a tariff structure $P^{\text{bench}}$ in place and a competitor introduces a tariff structure $P^{\text{comp}}$ based on a new modeling technique or a different set of rating variables. We define the relativity $r_i$ as the ratio of the competing premium to the benchmark premium for policyholder $i$:
\begin{equation}
r_i = \frac{P_i^{\text{comp}}}{P_i^{\text{bench}}}\,.
\label{relativities}
\end{equation}
A small relativity indicates a profitable policy which can potentially be lost to a competitor offering a lower premium. A high relativity reveals an underpriced policy which could benefit from better loss control measures such as renewal restrictions. These statements make the assumption that $P^{\text{comp}}$ is a more accurate reflection of the true risk compared to $P^{\text{bench}}$. 

\paragraph{Loss ratio lift} The loss ratio (LR) is the ratio of total incurred claim losses and total earned premiums. 
Following \citet{Goldburd2016}, we assess the loss ratio lift in the following way:
\vspace{-2.5mm}
\begin{enumerate}[itemsep=-0.1em]
	\item sort the policies from smallest to largest relativity $r_i$;
	\item bin the policies into groups containing the same amount of total exposure $e$;
	\item within each bin, calculate the overall loss ratio using the benchmark premium $P^{\text{bench}}$.
\end{enumerate}
\vspace{-2.5mm}
The bins should have loss ratios around \num{100}\% if the benchmark tariff is a technically accurate reflection of the risk. However, an upward trend in the loss ratios would indicate that policies with a lower (higher) premium under the competing tariff are those with a lower (higher) loss ratio in the benchmark tariff, pointing out that the competing tariff better aligns the risk. 

\paragraph{Double lift} A double lift chart facilitates a direct comparison between two potential tariff structures. Following \citet{Goldburd2016}, this chart is created in the following way:
\vspace{-2.5mm}
\begin{enumerate}[itemsep=-0.1em]
	\item sort the policies from smallest to largest relativity $r_i$;
	\item bin the policies into groups containing the same amount of total exposure $e$;
	\item within each bin, calculate the average actual loss amount ($L$) and the average predicted pure premium for both the models ($P^{\text{bench}}$ and $P^{\text{comp}}$);
	\item within each bin, calculate the percentage error for both models as $P/L - 1$.
\end{enumerate}
\vspace{-2.5mm}
The best tariff structure is the one with the percentage errors closest to zero, indicating that those premiums match the actual losses more closely.

\paragraph{Gini index} 
\cite{Frees2013} introduced the ordered Lorenz curve to compare the tariffs $P^{\text{bench}}$ and $P^{\text{comp}}$ by analyzing the distribution of losses versus premiums, where both are ordered by the relativities $r$ from Eq.~\eqref{relativities}. The ordered Lorenz curve is defined as follows:
\begin{equation*}
\left( \frac{\sum_{i=1}^n L_i  \, \mathbbm{1}\{F_{n}(r_i)\leq s\}}{\sum_{i=1}^n L_i} , \frac{\sum_{i=1}^n P^{\text{bench}}_i \,  \mathbbm{1}\{F_{n}(r_i)\leq s\}}{\sum_{i=1}^n P^{\text{bench}}_i} \right) \,,
\end{equation*}
for $s \in [0,1]$ where $F_{n}(r_i)$ is the empirical cumulative distribution function of the relativities $r$. This curve will coincide with the 45 degree line of equality when the technical pricing is done right by the benchmark premium. However, the curve will be concave up when $P^{\text{comp}}$ is able to spot tariff deficiencies in $P^{\text{bench}}$. The cumulative distributions are namely taken from the most overpriced policies towards the most underpriced policies in $P^{\text{bench}}$. The Gini index, introduced by \citet{Gini1912} and computed as twice the area between the ordered Lorenz curve and the line of equality, has a direct economic interpretation. A tariff structure $P^{\text{comp}}$ that yields a larger Gini index is likely to result in a more profitable portfolio because of better differentiation between good and bad risks. The insurer can decide to only retain the policies with a relativity value below a certain threshold. Averaging this decision over all possible thresholds, \citet{Frees2013} show that the average percentage profit for an insurer equals one half of the Gini index.

\subsection{Adverse selection and profits}
\label{adv_sel}
\begin{figure}[h!]
	\centering
	\includegraphics[width=\textwidth]{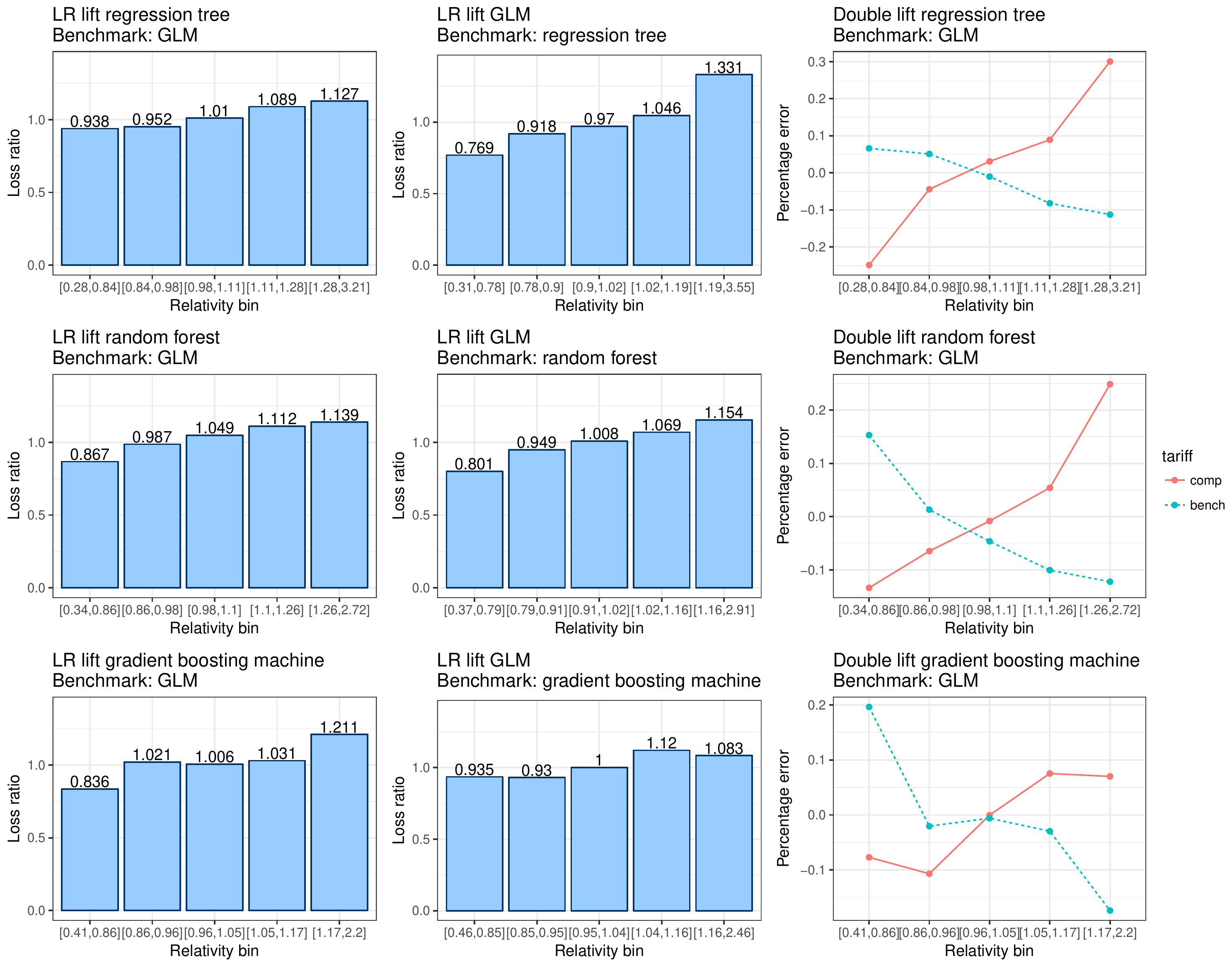}
	\caption{Assessment of model lift for the regression tree (top), random forest (middle) and gradient boosting machine (bottom). The left column shows the loss ratio lift for the tree-based techniques with the GLM as benchmark. The middle column shows the loss ratio lift for the GLM with the tree-based techniques as benchmark. The right column shows the double lift chart for the tree-based techniques with the GLM as benchmark.}
	\label{lift}
\end{figure}

The panels in the left column of Figure~\ref{lift} show the loss ratio lift charts for the regression tree, random forest and gradient boosting machine respectively with the GLM as benchmark tariff (i.e.,~the GLM premium is the denominator in Eq.~\eqref{relativities}). All tree-based methods show an increasing trend in the loss ratios. This implies that policies which would receive a lower premium under the competing tariff, those in the first bins, are policies with favorable loss ratios. At the same time, policies having a higher premium under the competing tariff, those in the last bins, exhibit detrimental loss ratios. The tree-based techniques are therefore able to spot deficiencies in the GLM benchmark tariff. One should not draw conclusions from these graphs too fast however. The middle panels of Figure~\ref{lift} show the loss ratio lifts for the GLM with the three respective tree-based techniques as a benchmark tariff. Comparing these lift charts side by side, we can observe that the upwards trend is now steeper in the cases of the regression tree and random forest. Thus, the GLM is better in spotting deficiencies in those tree-based tariffs compared to the other way around. The gradient boosting machine and GLM result in rather complementary tariffs. The GLM is very good in the three middle relativity bins, but the gradient boosting machine is clearly outperforming in the first and last bin.

These findings are confirmed by the double lift charts in the right panels of Figure~\ref{lift}. These show the double lift charts obtained with the GLM as the benchmark tariff in the relativities. The red and turquoise line respectively show the percentage error for the tree-based model and the GLM. For both the regression tree and the random forest, the percentage error for the GLM benchmark tariff is more closely centered around zero compared to the competitor percentage error. We again notice the complementarity of the gradient boosting machine and GLM tariffs. The percentage error for the gradient boosting machine is closer to zero for the first and last relativity bin, but the GLM is closer to zero for the other three relativity bins. 

The gradient boosting machine tariff clearly holds economic value over the GLM benchmark. However, in the bottom left panel of Figure~\ref{lift}, we observe that the gradient boosting machine is slightly over-correcting the GLM premium in the extreme ends of the tariff. The loss ratio in the first bin is \num{0.84} while the average relativity in that bin is equal to \num{0.78}. Likewise, the average loss ratio in the last bin is \num{1.21} while the average relativity in that bin is equal to \num{1.30}.

Table~\ref{gini_table} shows a two-way comparison of Gini indices for the machine learning methods and the GLM. The row names indicate the model generating the benchmark tariff structure $P^{\text{bench}}$ while the column names indicate the model generating the competing tariff structure $P^{\text{comp}}$. The row-wise maximum values are indicated in bold. We observe that the gradient boosting machine achieves the highest Gini index when the benchmark is either the GLM, the regression tree or the random forest. When the gradient boosting machine serves as benchmark, the GLM attains the highest Gini index. We use the mini-max strategy of \cite{Frees2013} where we search for the benchmark model with the minimal maximum Gini index. In other words, we look for the benchmark model with the lowest value in bold in Table~\ref{gini_table}. The gradient boosting machine achieves this minimal maximum Gini index, indicating that this approach leads to a tariff structure that is the least probable to suffer from adverse selection. Note that the GLM tariff achieves the second place, before the random forest and regression tree. 

\citet{Frees2013} explain that the average profit for an insurer is equal to half the Gini index. Let us assume that the insurance company uses state-of-the-art GLMs to develop their current tariff structure on this specific data. This implies that developing a competing tariff structure with gradient boosting machines would result in an average profit of around $3.3\%$ for the insurer. The average is taken over all possible decision-making strategies that the insurance company can take to retain policies based on the relativities. Therefore, by following an optimal strategy, the profit can even be higher for a specific choice of portfolio. We suspect that the improvement in profits could be even greater if there were more explanatory variables in the data.

\begin{table}[h!]
	\centering
	\begin{tabular}{llrrrr}
		\toprule
		& \multicolumn{1}{c}{competitors:} 
		& \multicolumn{1}{c}{GLM} 
		& \multicolumn{1}{c}{CART} 
		& \multicolumn{1}{c}{RF} 
		& \multicolumn{1}{c}{GBM} \\
		\midrule
		\multirow{4}{*}{\rotatebox[origin=c]{90}{benchmark}} & GLM &  & 4.07 & 5.99 & \textbf{6.57} \\
		& CART & 10.86 &  & 10.10 & \textbf{12.02}\\
		& RF & 7.07 & 0.53  & & \textbf{7.59}  \\
		& GBM & \textbf{3.93} & 0.67 & 2.30  & \\
		\bottomrule
	\end{tabular}
	\caption{Two-way comparison of Gini indices for the different tree-based techniques and GLM.}
	\label{gini_table}
\end{table}

\subsection{Solidarity and risk differentiation}

\begin{figure}[h!]
	\centering
	\includegraphics[width=0.94\textwidth]{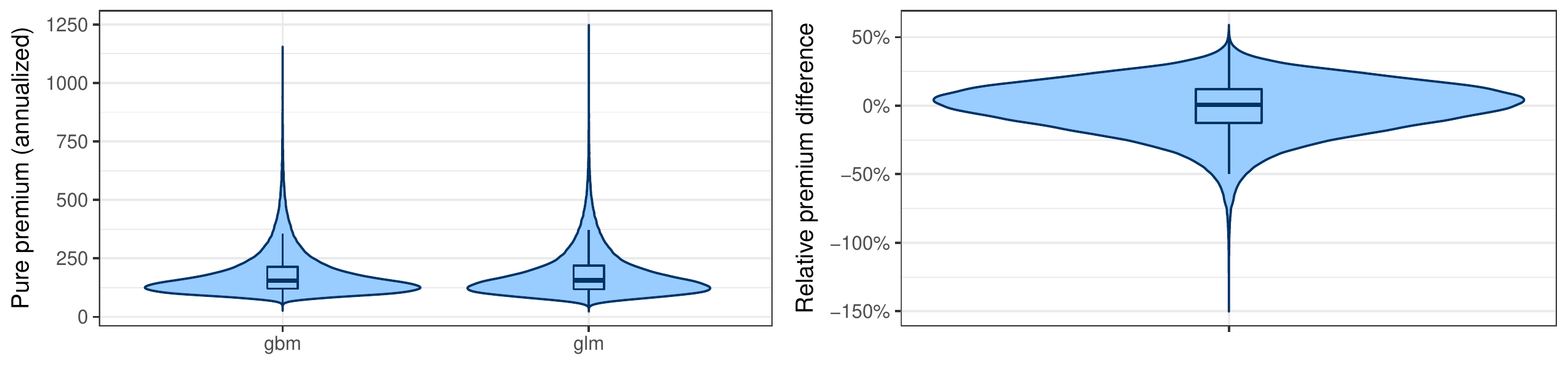}
	\caption{Comparison of the annual premium distribution in the gradient boosting machine tariff $P^{\text{gbm}}$ and the GLM tariff $P^{\text{glm}}$: absolute premiums (left) and relative differences (right).}
	\label{prem_dist}
\end{figure}

From a social point of view, it is crucial for everybody to be able to buy insurance cover at a reasonable price. A tariff structure that follows from a complex machine learning algorithm should not lead to the ``personalization of risk'' with excessively high premiums for some policyholders. Figure~\ref{prem_dist} shows violin plots of the annual (i.e.,~exposure equals one) premium distribution in both the gradient boosting machine tariff $P^{\text{gbm}}$ and the GLM tariff $P^{\text{glm}}$. We only consider the gradient boosting machine because Sections \ref{oos_comp} and \ref{adv_sel} teach us that only this method holds added value over the GLM. The left panel shows the annual premium amounts and we observe that both distributions look very similar. The minimum, median and maximum premium is 43, 155 and 1138 Euro in $P^{\text{gbm}}$ and 41, 156 and 1230 Euro in $P^{\text{glm}}$ respectively. The right panel shows the relative difference between both premiums, namely $(P^{\text{gbm}} - P^{\text{glm}})/P^{\text{gbm}}$. The difference is centered around zero and for half the policyholders the difference lies in the range $[-12\%,+12\%]$. This implies that, overall, $P^{\text{gbm}}$ and $P^{\text{glm}}$ trade off segmentation and risk pooling in a similar way, thereby finding a balance between differentiation and solidarity. For a small selection of policyholders, the gradient boosting machine leads to considerable discounts compared to the GLM.

\begin{figure}[h!]
	\centering
	\includegraphics[width=0.94\textwidth]{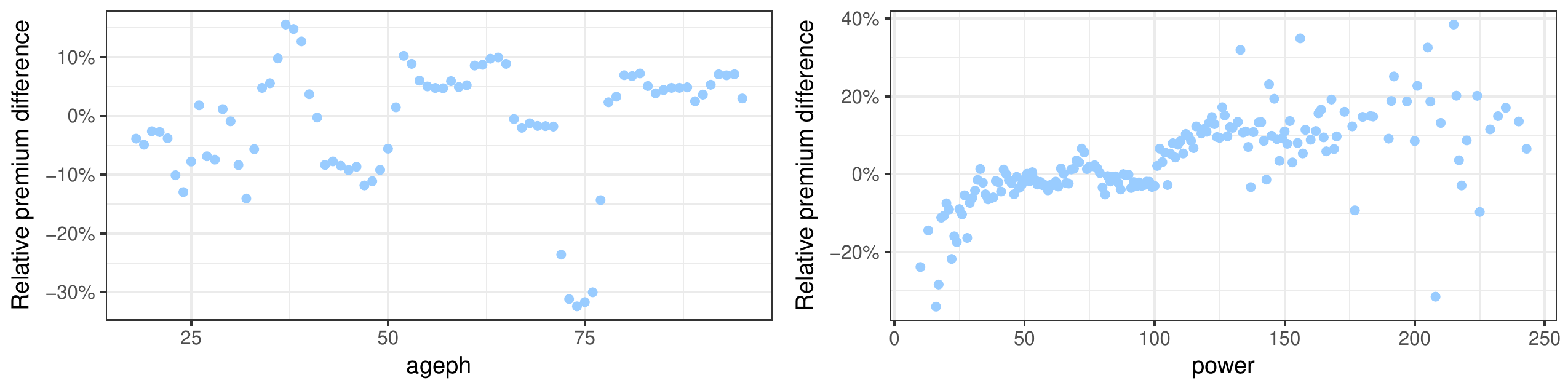}
	\caption{Comparison of premium differences between $P^{\text{gbm}}$ and $P^{\text{glm}}$ over the age of the policyholder (left) and the power of the car (right).}
	\label{prem_var}
\end{figure}

The gradient boosting machine and GLM result in similar premiums on a portfolio level (see Table~\ref{prem_vs_loss}), but on a coarser scale, they could lead to different approaches for targeting specific customer segments. Figure~\ref{prem_var} are the relative premium differences between $P^{\text{gbm}}$ and $P^{\text{glm}}$ over the age of the policyholder in the left panel and the power of the car in the right panel. The blue dots show the average premium difference for policyholders with that specific characteristic, e.g.,~all policyholders aged 25. We observe that younger policyholders obtain a slightly lower premium in the gradient boosting machine tariff, while senior policyholders obtain a slightly higher premium compared to the GLM tariff. For middle aged policyholders there are some fluctuations which can be explained by analyzing the age effects in Figure~\ref{pdp_freq_ageph}. For the group of dots around the age of 75, $P^\text{gbm}$ gives an average 30\% premium discount over $P^\text{glm}$. Figure~\ref{pdp_freq_ageph} shows that the age effect starts increasing before the age of 75 in the GLM (top left), but only after the age of 75 for the gradient boosting machine (bottom right). Therefore, policyholders around the age of 75 obtain a better deal in the gradient boosting machine tariff. In the right panel of Figure~\ref{prem_var}, the premium differences increase roughly monotonically with the power of the car. Low powered cars obtain a lower premium in $P^{\text{gbm}}$ while high powered cars get a lower premium in $P^{\text{glm}}$. In between the differences are close to zero, indicating that both tariffs treat those cars in a similar fashion.

We conclude that gradient boosting machines can be a valuable tool for the insurer, while the other tree-based techniques under investigation show little added value on our specific portfolio. The gradient boosting machine is able to deliver a tariff that assesses the underlying risk in a more accurate way, thereby guarding the insurer against adverse selection risks which eventually can result in a profit. The gradient boosting machine also honored the principle of solidarity in the portfolio, offering affordable insurance cover for all policyholders with premiums in the same range as the benchmark GLM.

\section{Conclusions and outlook}
\label{conclusion}
In this study, we have adapted tree-based machine learning to the problem of insurance pricing, thereby leaving the comfort zone of both traditional ratemaking and machine learning. State-of-the-art GLMs are compared to regression trees, random forests and gradient boosting machines. These tree-based techniques can be used on insurance data, but care has to be taken with the underlying statistical assumptions in the form of the loss function choice. This paper brings multiple contributions to the existing literature. First, we develop complete tariff plans with tree-based machine learning techniques for a real-life insurance portfolio. In this process, we use the Poisson and gamma deviance because the classical squared error loss is not appropriate for a frequency-severity problem. Second, our elaborate cross-validation scheme gives a well thought and careful tuning procedure, allowing us to assess not only the performance of different methods, but also the stability of our results across multiple data folds. Third, we go beyond a purely statistical comparison and also focus on business metrics used in insurance companies to evaluate different tariff strategies. Fourth, we spend a lot of attention on the interpretability of the resulting models. This is a very important consideration for insurance companies within the GDPR's regime of algorithmic accountability. Fifth, our complete analysis is available in well-documented \textsf{R} functions, readily applicable to other data sets. This includes functions for training, predicting and evaluating models, running the elaborate cross-validation scheme, interpreting the resulting models and assessing the economic lift of these models. Sixth, we extended the \texttt{rpart} package such that it is now possible to build regression trees with a gamma deviance as loss function and random forest with both the Poisson and gamma deviance as loss functions. This package is available at \url{https://github.com/henckr/distRforest}. 

The gradient boosting machine is consistently selected as best modeling approach, both by out-of-sample performance measures and model lift assessment criteria. This implies that an insurer can prevent adverse selection and generate profits by considering this new modeling framework. However, this might be impossible because of regulatory issues, e.g.,~filing requirements (see Appendix~\ref{App_filreq}). In that case, an insurance company can still learn valuable information on how to form profitable portfolios from an internal, technical model and translate this to a commercial product which is in line with all those requirements. A possible approach would be to approximate a gradient boosting machine with a GLM, much in line with the strategy to develop the benchmark pricing GLM in this study. The gradient boosting machine can be used to discover the important variables and interactions between those variables, which can then be included in a GLM for deployment. Although we present the tools to detect potentially interesting variables and interactions, we leave for future work the building of a competitive GLM inspired by the gradient boosting machine.



\section*{Acknowledgments}

This research was supported in part by the Research Foundation - Flanders (FWO) [SB grant 1S06018N] and by the Society of Actuaries James C. Hickman Scholar Program. Furthermore, Katrien Antonio acknowledges financial support from the Ageas Research Chair at KU Leuven and from KU Leuven's research council (project COMPACT C24/15/001). We are grateful for the access to the computing resources of the Department of Mathematics and Statistics at McGill University and the VSC (Flemish Supercomputer Center), funded by the Research Foundation - Flanders (FWO) and the Flemish Government - department EWI. Thanks are also due to Christian Genest for supporting this project in various ways and to Michel Denuit for sharing the data analyzed in this paper. Finally, we would like to thank the editor and reviewers for their comments which substantially improved the paper.

{\small
\bibliography{bibfile}}

\newpage
\appendix

\section{List of variables in the MTPL data}
\label{App_var}

\begin{table}[h!]
	\centering
	\begin{tabular}{lp{14cm}}
		\toprule
		\multicolumn{2}{l}{\textbf{Claim information and exposure-to-risk measure}} \\
		\texttt{nclaims} & The number of claims filed by the policyholder. \\
		\texttt{amount} & The total amount claimed by the policyholder in euro.\\
		\texttt{expo} & The fraction of the year during which the insurer was exposed to the risk. \\
		\midrule
		\multicolumn{2}{l}{\textbf{Categorical risk factors}} \\
		\texttt{coverage} & Type of coverage provided by the insurance policy: TPL, TPL+ or TPL++. \\
		& TPL = only third party liability, \\
		& TPL+ = TPL + limited material damage, \\
		& TPL++ = TPL + comprehensive material damage. \\
		\texttt{fuel} & Type of fuel of the vehicle: gasoline or diesel. \\
		\texttt{sex} & Gender of the policyholder: male or female. \\
		& {\scriptsize (As from 21 December 2012, the European Court of Justice prohibited the use of gender in insurance tariffs to avoid discrimination between males and females, known as the Test-Achats Ruling. Gender is therefore only investigated for use within an internal technical tariff, but can not be used in a commercial product.)} \\
		\texttt{use} & Main use of the vehicle: private or work. \\
		\texttt{fleet} & The vehicle is part of a fleet: yes or no. \\
		\midrule
		\multicolumn{2}{l}{\textbf{Continuous risk factors}} \\
		\texttt{ageph} & Age of the policyholder in years. \\
		\texttt{power} & Horsepower of the vehicle in kilowatt. \\
		\texttt{agec} & Age of the vehicle in years. \\
		\texttt{bm} & Level occupied in the former compulsory Belgian bonus-malus scale. \\
		& From 0 to 22, a higher level indicates a worse claim history, see \citet{Lemaire1995}. \\
		& {\scriptsize (This variable is typically not used as an a priori rating factor, but rather as an a posteriori correction in a credibility framework or bonus-malus scheme. We however keep \texttt{bm} in the data to assess the amount of information contained in this variable and to investigate the resulting effect.)} \\
		\midrule
		\multicolumn{2}{l}{\textbf{Spatial risk factor}} \\
		\texttt{long} & Longitude coordinate of the center of the municipality where the policyholder resides. \\
		\texttt{lat} & Latitude coordinate of the center of the municipality where the policyholder resides. \\
		\bottomrule
	\end{tabular}
	\caption{Description of the available variables in the \texttt{MTPL} data.}
	\label{variables}
\end{table}

\newpage

\section{Search grid for the tuning parameters}
\label{App_grid}
\begingroup
\renewcommand{\arraystretch}{1}
\begin{table}[h]
	\centering
	\begin{tabular}{l@{\hskip 1cm}c}
		\toprule
		\multirow{2}{*}{Regression tree}  &  $cp \in \{\num{1.0e-05}, \num{1.1e-05}, \ldots, \num{1.0e-02}\}$   \\
		& $\gamma \in \{2^{-6}, 2^{-5}, 2^{-4}, 2^{-3}, 2^{-2}, 2^{-1}, 2^{0}\}^{\boldsymbol{*}}$  \\
		\midrule
		\multirow{2}{*}{Random forest}  & $T \in \{\num{100}, \num{200}, \ldots, \num{5000}\}$  \\
		& $m \in \{1, 2, 3, 4, 5, 6, 7, 8, 9, 10, 11\}$  \\
		\midrule
		\multirow{2}{*}{Gradient boosting machine} & $T \in \{\num{100}, \num{200}, \ldots, \num{5000}\}$  \\
		& $d \in \{1, 2, 3, 4, 5, 6, 7, 8, 9, 10\}$  \\
		\bottomrule
	\end{tabular}
	\caption{Search grid for the tuning parameters in the different tee-based machine learning techniques. \\ $^{\boldsymbol{*}}$ Note that the $\gamma$ tuning parameter is only used in frequency models for the Poisson deviance.}
	\label{grid}
\end{table}
\endgroup



\section{Random forests for claim frequency data}
\label{rf_extra}
\begin{figure}[h!]
	\centering
	\includegraphics[width=\textwidth]{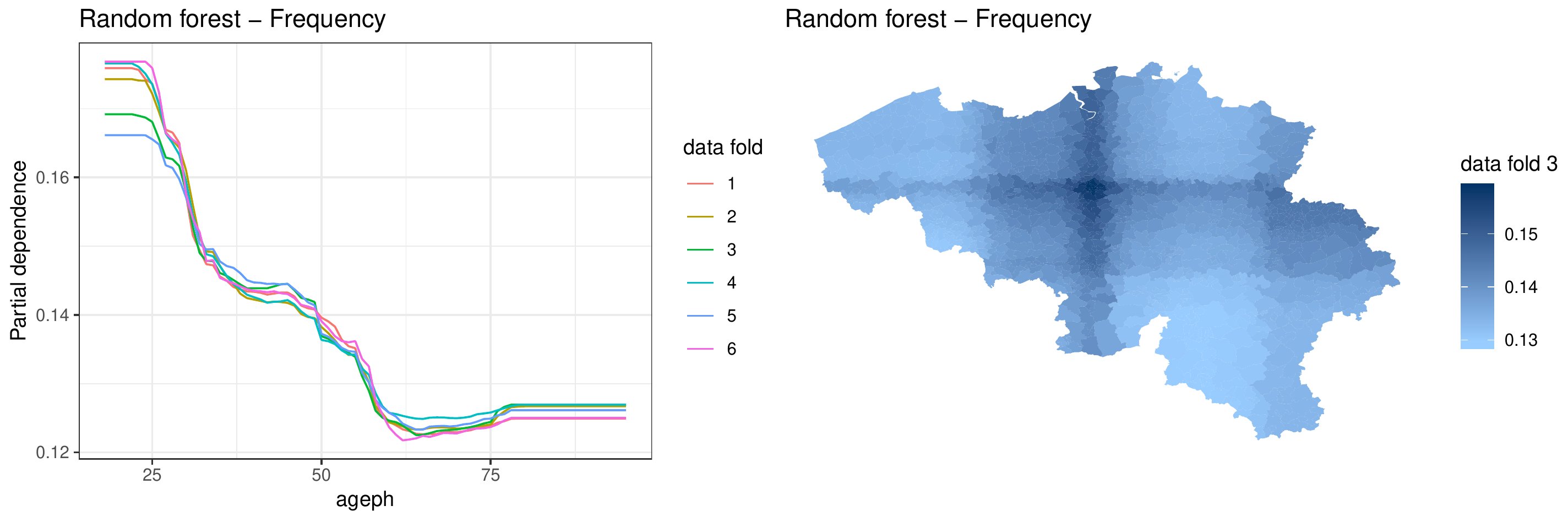}
	\caption{Effect of the age of the policyholder (left) and the municipality of residence (right) on frequency in a random forest.}
	\label{pdp_freq_rf}
\end{figure}

The left and right panel of Figure~\ref{pdp_freq_rf} show the partial dependence plot of the age and spatial effect in the claim frequency random forests, respectively. These effects are discussed in Section~\ref{modinterpret}. The six random forests for frequency contain rather different number of trees, ranging from \num{100} to \num{5000} (see Table~\ref{opt_params} earlier). However, these random forests exhibit very similar effects for the age of the policyholder in Figure~\ref{pdp_freq_rf}. This indicates that the underlying model structure does not change drastically after including a sufficient number of trees.

\newpage

\section{Filing requirements for pricing models}
\label{App_filreq}
Insurance companies can be required by regulation to file their rating model on paper. This section presents such frequency and severity models, trained on the data where $\mathcal{D}_3$ was kept as hold-out test set. Section~\ref{App_glm} and \ref{App_tree} show the GLMs and regression trees respectively. The other five GLMs and regression trees are not shown due to lack of space. This filing requirement is more difficult to satisfy for the ensemble methods, but it is possible by printing the individual trees. However, this would result in a large amount of pages which is not very practical or insightful for the regulator.

\subsection{GLM}
\label{App_glm}

\begin{table}[h!]  \scriptsize
	\begin{subtable}{0.5\textwidth}
		\setlength\tabcolsep{2pt}
		\begin{tabular}{lSSSS} 
			\toprule
			& \multicolumn{1}{c}{Coefficient} & \multicolumn{1}{c}{SE ($\sigma$)} & \multicolumn{1}{c}{$t$-stat} & \multicolumn{1}{r}{$p$-value}\\ 
			\midrule
			Intercept & -2.19$^{***}$ &  0.03   &  -64.09 &  0.00 \\ 
			coverageTPL+ & -0.10$^{***}$ &  0.02   &  -5.46	&  0.00 \\ 
			coverageTPL++ & -0.10$^{***}$	&  0.03  	&  -3.97	&  0.00 \\ 
			fueldiesel & 0.18\,\,$^{***}$ &  0.02  	&  10.00	&  0.00 \\ 
			fleetyes & -0.13$^{***}$	&  0.05  	&  -2.72	&  0.01 \\ 
			ageph[18,26) & 0.32\,\,$^{***}$ 	&  0.04   	&  8.94 	&  0.00 \\ 
			ageph[26,29) & 0.16\,\,$^{***}$ 	&  0.04   	&  4.61 	&  0.00 \\ 
			ageph[29,32) & 0.11\,\,$^{***}$ 	&  0.04   	&  3.11	&  0.00 \\ 
			ageph[32,35) & 0.02 	&  0.04   	&  0.46	&  0.65 \\ 
			ageph[50,54) & -0.11$^{***}$	&  0.03  	&  -3.53	&  0.00 \\ 
			ageph[54,58) & -0.05	&  0.04   	&  -1.28	&  0.20 \\ 
			ageph[58,62) & -0.23$^{***}$	&  0.05  	&  -4.93	&  0.00 \\ 
			ageph[62,73) & -0.25$^{***}$ &  0.04  	&  -7.06	&  0.00 \\ 
			ageph[73,95] & -0.21$^{***}$	&  0.05  	&  -4.42	&  0.00 \\ 
			bm[1,2) & 0.14\,\,$^{***}$ 	&  0.03   	&  5.46 	&  0.00 \\ 
			bm[2,3) & 0.16\,\,$^{***}$ 	&  0.04   	&  4.48 	&  0.00 \\ 
			bm[3,5) & 0.37\,\,$^{***}$ 	&  0.03  	&  11.86 	&  0.00 \\ 
			bm[5,7) & 0.32\,\,$^{***}$	&  0.03  	&  11.39	&  0.00 \\ 
			bm[7,9) & 0.48\,\,$^{***}$ 	&  0.03   	&  14.77 	&  0.00 \\ 
			bm[9,11) & 0.52\,\,$^{***}$ 	&  0.03   	&  17.30 	&  0.00 \\ 
			bm[11,22] & 0.75\,\,$^{***}$ 	&  0.03   	&  26.07 	&  0.00 \\ 
			agec[9,14) & 0.04\,\,$^{*}$ 	&  0.02   	&  1.91 	&  0.06 \\ 
			agec[14,48] & -0.02	&  0.03  	&  -0.75	&  0.46 \\ 
			\midrule
			\textit{Significance codes:}  & \multicolumn{4}{l}{$^{*}p<0.1$; $^{**}p<0.05$; $^{***}p<0.01$} \\ 
			\bottomrule
		\end{tabular}
	\end{subtable}%
	\begin{subtable}{0.5\textwidth}
		\setlength\tabcolsep{2pt}
		\begin{tabular}{lSSSS} 
			\toprule
			& \multicolumn{1}{c}{Coefficient} & \multicolumn{1}{c}{SE ($\sigma$)} & \multicolumn{1}{c}{$t$-stat} & \multicolumn{1}{r}{$p$-value}\\ 
			\midrule
			power[10,35) & -0.16$^{***}$ 	&  0.04   	&  -3.95	&  0.00 \\ 
			power[35,42) & -0.05	&  0.03  	&  -1.54 	&  0.12 \\ 
			power[42,49) & -0.02	&  0.03   	&  -0.55	&  0.58 \\ 
			power[59,73) & 0.02	&  0.02   	&  0.89 	&  0.37 \\ 
			power[73,92) & 0.07\,\,$^{*}$ 	&  0.04   	&  1.76 	&  0.08 \\ 
			power[92,243] & 0.13\,\,$^{***}$	&  0.05   	&  2.88 	&  0.00 \\ 
			latlong[-0.46,-0.36)   & -0.31 	&  0.19   	&  -1.62 	&  0.11 \\ 
			latlong[-0.36,-0.26)   & -0.33$^{***}$ 	&  0.06   	&  -5.97 	&  0.00 \\ 
			latlong[-0.26,-0.18)   & -0.20$^{***}$ 	&  0.03   	&  -5.86 	&  0.00 \\ 
			latlong[-0.18,-0.12)   & -0.23$^{***}$ 	&  0.04   	&  -6.38 	&  0.00 \\ 
			latlong[-0.12,-0.061)   & -0.13$^{***}$ 	&  0.03   	&  -4.41 	&  0.00 \\ 
			latlong[-0.061,-0.017)   & -0.05 	&  0.03   	&  -1.62 	&  0.11 \\ 
			latlong[0.025,0.077)   & 0.06\,\,$^{**}$ 	&  0.03  	&  2.42 	&  0.02 \\ 
			latlong[0.077,0.14)   & 0.03 	&  0.03   	&  0.86 	&  0.39 \\ 
			latlong[0.14,0.23)   & 0.04 	&  0.03   	&  1.29 	&  0.20 \\ 
			latlong[0.23,0.33] & 0.36\,\,$^{***}$ 	&  0.03   	&  12.37 	&  0.00 \\ 
			agephpower-0.05 & -0.04 	&  0.06   	&  -0.61	&  0.54 \\ 
			agephpower-0.02 & -0.11$^{***}$ 	&  0.04   	&  -3.15 	&  0.00 \\ 
			agephpower-0.01 & -0.03 	&  0.03   	&  -0.79 	&  0.43 \\ 
			agephpower0.01 & -0.00 	&  0.03   	&  -0.02 	&  0.98 \\ 
			agephpower0.02 & -0.04 	&  0.05   	&  -0.67 	&  0.50 \\ 
			agephpower0.04 & 0.04 &  0.04   	&  1.07 	&  0.29 \\ 
			&&&&\\
			\midrule
			\textit{Significance codes:}  & \multicolumn{4}{l}{$^{*}p<0.1$; $^{**}p<0.05$; $^{***}p<0.01$} \\ 
			\bottomrule
		\end{tabular} 
	\end{subtable}
	\caption{Frequency GLM specification, trained on the data where $\mathcal{D}_3$ was kept as hold-out test set.} 
	\label{glm_freq_data3} 
\end{table}

\begin{table}[h!]  \scriptsize
	\begin{subtable}{0.5\textwidth}
		\setlength\tabcolsep{2pt}
		\begin{tabular}{lSSSS} 
			\toprule
			& \multicolumn{1}{c}{Coefficient} & \multicolumn{1}{c}{SE ($\sigma$)} & \multicolumn{1}{c}{$t$-stat} & \multicolumn{1}{r}{$p$-value}\\ 
			\midrule
			Intercept & 7.31\,\,$^{***}$ 		&  0.08  		&  88.50 &  0.00 \\ 
			coverageTPL+ & -0.23$^{***}$ 		&  0.05   		&  -4.62 	&  0.00 \\ 
			coverageTPL++ & 0.18\,\,$^{**}$ 		&  0.07   		&  2.54 		&  0.01 \\ 
			ageph[18,25) & 0.13 		&  0.11   		&  1.23 		&  0.22 \\ 
			ageph[25,28) & -0.02 		&  0.10   		&  -0.15 		&  0.88 \\ 
			ageph[28,30) & 0.08 		&  0.11   		&  0.70 		&  0.49 \\ 
			ageph[30,33) & 0.07 		&  0.10   		&  0.65		&  0.52 \\ 
			ageph[33,36) & -0.19$^{*}$ 		&  0.10   		&  -1.87 		&  0.06 \\ 
			ageph[36,39) & -0.26$^{**}$ 		&  0.10   		&  -2.57 		&  0.01 \\ 
			ageph[39,42) & -0.19$^{*}$ 		&  0.10   		&  -1.84 		&  0.07 \\ 
			ageph[42,45) & -0.01 		&  0.10   		&  -0.12 		&  0.91 \\ 
			ageph[49,52) & -0.02 		&  0.10   		&  -0.21 		&  0.83 \\ 
			ageph[52,55) & -0.13 		&  0.11   		&  -1.18 		&  0.24 \\ 
			ageph[55,61) & -0.17$^{*}$ 		&  0.10   		&  -1.69 		&  0.09 \\ 
			\midrule
			\textit{Significance codes:}  & \multicolumn{4}{l}{$^{*}p<0.1$; $^{**}p<0.05$; $^{***}p<0.01$} \\ 
			\bottomrule
		\end{tabular}
	\end{subtable}%
	\begin{subtable}{0.5\textwidth}
		\setlength\tabcolsep{2pt}
		\begin{tabular}{lSSSS} 
			\toprule
			& \multicolumn{1}{c}{Coefficient} & \multicolumn{1}{c}{SE ($\sigma$)} & \multicolumn{1}{c}{$t$-stat} & \multicolumn{1}{r}{$p$-value}\\ 
			\midrule
			ageph[61,66) & -0.20$^{*}$ 		&  0.11   		&  -1.73 		&  0.08 \\ 
			ageph[66,72) & -0.04 		&  0.11  		&  -0.33 		&  0.75 \\ 
			ageph[72,95] & 0.11 		&  0.11   		&  0.93 		&  0.35 \\ 
			agec[0,2) & 0.22\,\,$^{**}$ 		&  0.11   		&  2.09 		&  0.04 \\ 
			agec[2,3) & -0.11 		&  0.09   		&  -1.11 		&  0.27 \\ 
			agec[3,4) & -0.08 		&  0.10   		&  -0.82 		&  0.41 \\ 
			agec[4,6) & -0.08 		&  0.07   		&  -1.09 		&  0.28 \\ 
			agec[6,7) & -0.10		&  0.08   		&  -1.22 		&  0.22 \\ 
			agec[7,8) & -0.10 		&  0.09   		&  -1.21 		&  0.23 \\ 
			agec[10,11) & -0.12 		&  0.09   		&  -1.33 		&  0.18 \\ 
			agec[11,12) & -0.05 		&  0.09   		&  -0.52 		&  0.60 \\ 
			agec[12,14) & -0.14$^{*}$ 		&  0.08   		&  -1.69 		&  0.09 \\ 
			agec[14,48] & -0.01 		&  0.09   		&  -0.06 		&  0.95 \\ 
			&&&&\\
			\midrule
			\textit{Significance codes:}  & \multicolumn{4}{l}{$^{*}p<0.1$; $^{**}p<0.05$; $^{***}p<0.01$} \\ 
			\bottomrule
		\end{tabular} 
	\end{subtable}
	\caption{Severity GLM specification, trained on the data where $\mathcal{D}_3$ was kept as hold-out test set.} 
	\label{glm_sev_data3} 
\end{table}

\subsection{Regression tree}
\label{App_tree}

\begin{figure}[h!] \centering
	\begin{tikzpicture}[      
	every node/.style={anchor=south east,inner sep=0pt},
	x=1mm, y=1mm,
	]   
	\node (fig1) at (0,0)
	{\includegraphics[scale=0.6]{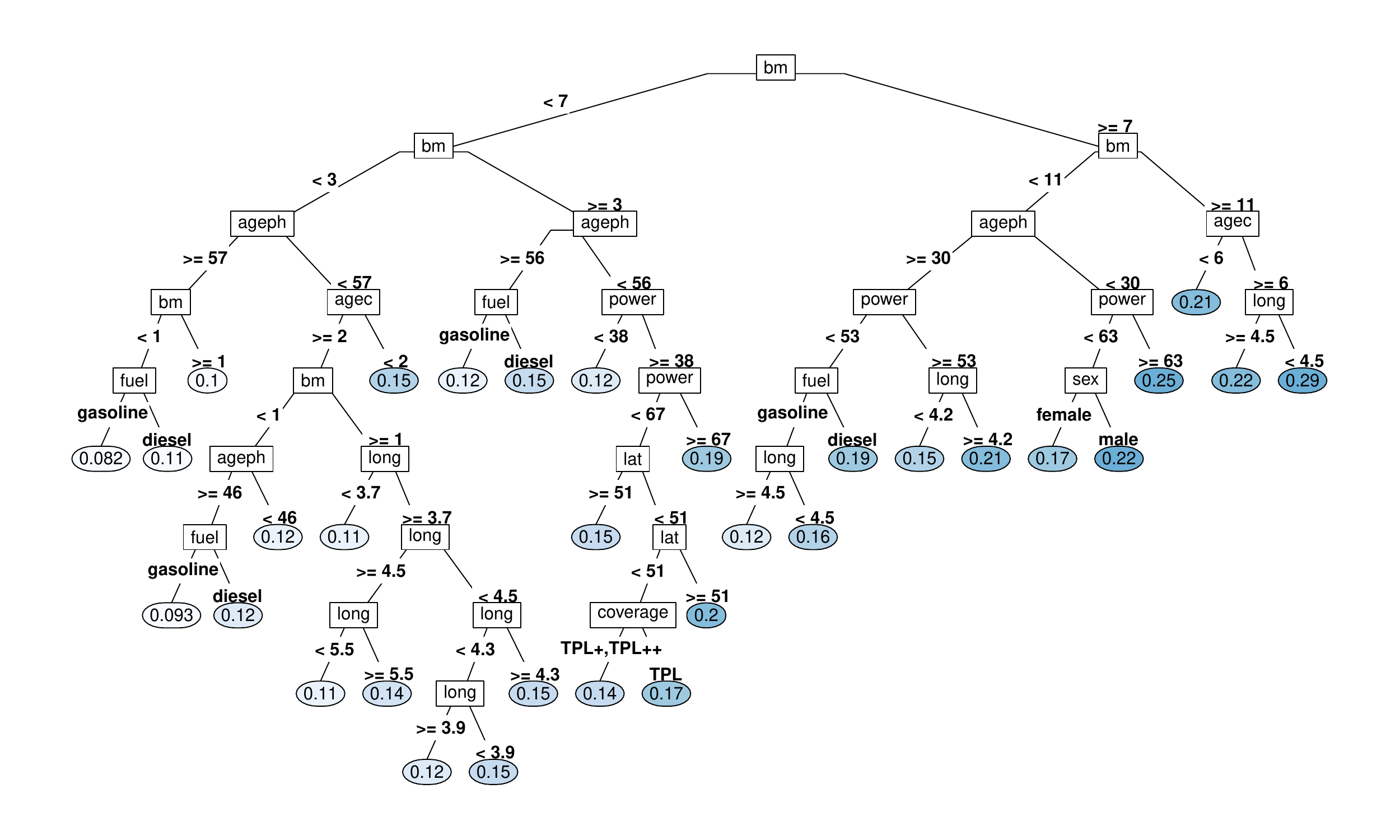}};
	\node (fig2) at (0,6)
	{\includegraphics[scale=0.6]{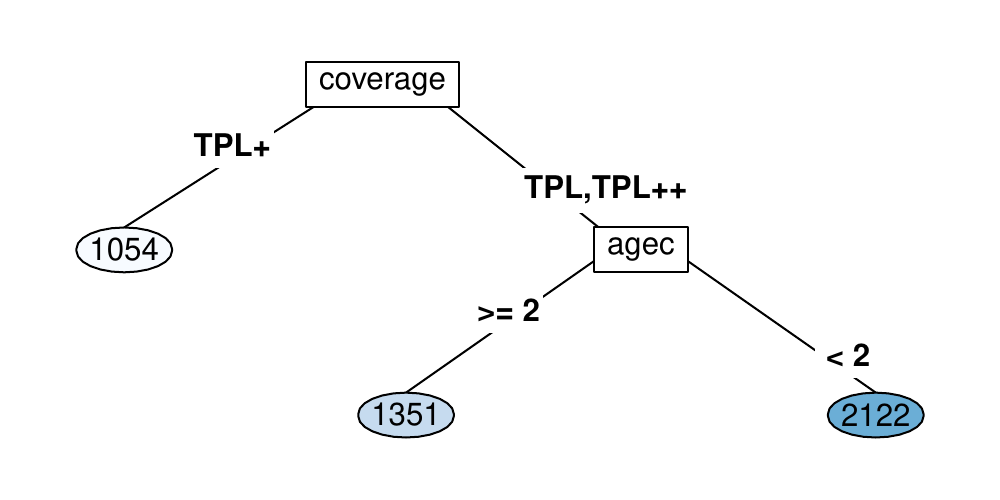}};  
	\end{tikzpicture}
	\caption{Regression trees for claim frequency (big, top left) and severity (small, bottom right), both trained on the data where $\mathcal{D}_3$ was kept as hold-out test set.}
\end{figure}

\section{Supplementary material}
\label{App_supp}
Supplementary material related to this article can be found online at \url{https://github.com/henckr/sevtree}.

\end{document}